\documentclass[longbibliography, aps, prx, amsmath, amssymb, 
amsfonts, twocolumn,superscriptaddress, footinbib]{revtex4-2}
\usepackage[german,american,english]{babel}
\usepackage{graphicx}
\usepackage{subcaption}
\usepackage{dcolumn}
\usepackage{bm}
\usepackage{amssymb}
\usepackage{amsmath}
\usepackage{amsfonts}
\usepackage{epsfig}
\usepackage{mathbbol}
\usepackage{latexsym}
\usepackage{theorem}
\usepackage{dcolumn}
\usepackage{algorithm}
\usepackage{algpseudocode}
\usepackage{hyperref}
\usepackage{float}
\usepackage[normalem]{ulem}
\usepackage{xcolor}
\usepackage{epstopdf}
\usepackage{physics}
\usepackage{lipsum}
\usepackage{microtype}
\hypersetup{                    
	colorlinks,
	citecolor=blue,
	filecolor=blue,
	linkcolor=blue,
	urlcolor=blue
}
\usepackage[utf8]{inputenc}
\UseRawInputEncoding
\begin{document}
	\title{Unified Quantum State Tomography and Hamiltonian Learning Using Transformer Models: A Language-Translation-Like Approach for Quantum Systems}
	
	\author{Zheng An}
	\affiliation{Department of Physics, The Hong Kong University of Science and Technology, Clear Water Bay, Kowloon, Hong Kong, China}
	
	\author{Jiahui Wu}
		\affiliation{Department of Physics, The Hong Kong University of Science and Technology, Clear Water Bay, Kowloon, Hong Kong, China}
	
	\author{Muchun Yang}
	   \affiliation{Institute of Physics, Beijing National Laboratory for Condensed
		Matter Physics,\\Chinese Academy of Sciences, Beijing 100190, China}
	\affiliation{School of Physical Sciences, University of Chinese Academy of
		Sciences, Beijing 100049, China}
	
	\author{D. L. Zhou} 
	\email{zhoudl72@iphy.ac.cn}
	\affiliation{Institute of Physics, Beijing National Laboratory for Condensed
	Matter Physics,\\Chinese Academy of Sciences, Beijing 100190, China}
	\affiliation{School of Physical Sciences, University of Chinese Academy of
	Sciences, Beijing 100049, China} \affiliation{Collaborative Innovation Center
	of Quantum Matter, Beijing 100190, China}
	\affiliation{Songshan Lake Materials Laboratory, Dongguan, Guangdong 523808,
	China}

	\author{Bei Zeng}%
	\email{zengb@ust.hk}
	\affiliation{Department of Physics, The Hong Kong University of Science and Technology, Clear Water Bay, Kowloon, Hong Kong, China}
	
	\date{\today} 
	
	\begin{abstract}
Schr{\"o}dinger's equation serves as a fundamental component in characterizing quantum systems, wherein both quantum state tomography and Hamiltonian learning are instrumental in comprehending and interpreting quantum systems. While numerous techniques exist for carrying out state tomography and learning Hamiltonians individually, no method has been developed to combine these two aspects. In this study, we introduce a new approach that employs the attention mechanism in transformer models to effectively merge quantum state tomography and Hamiltonian learning. By carefully choosing and preparing the training data, our method integrates both tasks without altering the model's architecture, allowing the model to effectively learn the intricate relationships between quantum states and Hamiltonian. We also demonstrate the effectiveness of our approach across various quantum systems, ranging from simple 2-qubit cases to more involved 2D antiferromagnetic Heisenberg structures. The data collection process is streamlined, as it only necessitates a one-way generation process beginning with state tomography. Furthermore, the scalability and few-shot learning capabilities of our method could potentially minimize the resources required for characterizing and optimizing quantum systems. Our research provides valuable insights into the relationship between Hamiltonian structure and quantum system behavior, fostering opportunities for additional studies on quantum systems and the advancement of quantum computation and associated technologies.
	\end{abstract}

	\maketitle
	
	\section{Introduction}
\begin{figure*}
		\centering
		\includegraphics[width=0.75\textwidth]{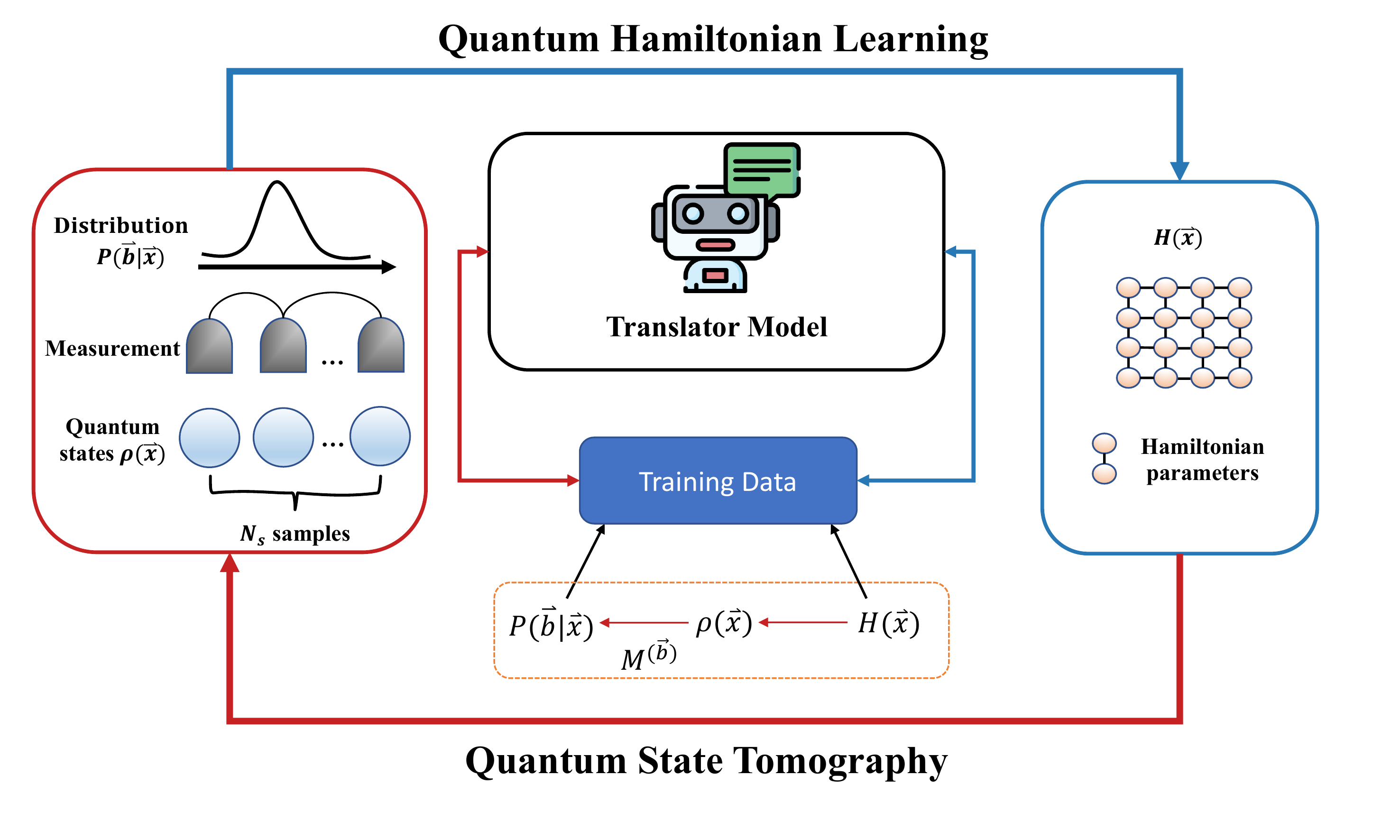}
		\caption{Bidirectional Translation Model for Quantum State Tomography and Hamiltonian Learning: Our model functions as an adaptable and efficient mediator between Quantum State Tomography and Hamiltonian Learning, effectively facilitating the interplay between the elucidation of quantum states derived from Hamiltonian parameters and the estimation of Hamiltonian parameters informed by observed ground states. The training data necessitates unidirectional generation from state tomography $H(\vec{x})\rightarrow P(\vec{b}|\vec{x})$, proving to be advantageous for the concurrent training of both methodologies.}
		\label{fig:translation model}
	\end{figure*}

	Quantum systems are governed by the Schr{\"o}dinger equation, which plays a pivotal role in defining the relationship between the Hamiltonian structure and the states of the system. This relationship is central to understanding the behavior of quantum systems~\cite{leimkuhler2004simulating} and for applications such as quantum computing and communication~\cite{nielsen2010quantum,preskill2018quantum}. 
	Moreover, the mapping between the Hamiltonian and the quantum states of a system is indispensable in quantum information science, as it enables us to predict the system's behavior~\cite{RevModPhys.89.035002,PhysRevLett.105.150401,Wang2017,Granade2012,PhysRevA.90.022117}. This knowledge is crucial in quantum computing applications, where Hamiltonian parameters are utilized to control and manipulate quantum systems for specific tasks~\cite{PhysRevResearch.1.033092,PhysRevA.65.042301,Handel2005,cao2022quantum,an2021learning}. Research in this domain can be bifurcated into two primary directions: Quantum State Tomography (QST) and Hamiltonian learning (see Fig.~\ref{fig:translation model}).
	
	Quantum state tomography and Hamiltonian learning are two fundamental techniques in the field of quantum information science, each with their own strengths and weaknesses. QST is a powerful method for characterizing quantum states comprehensively~\cite{d2003quantum,gross2010quantum,roos2004bell,vogel1989determination,PhysRevLett.118.020401}, though it can be computationally intensive, particularly for large and complex systems~\cite{haffner2005scalable,lu2007experimental}. Machine learning has been employed to improve accuracy and efficiency in QST~\cite{carrasquilla2019reconstructing,Torlai2018,PhysRevLett.123.230504,Xin2019,Lohani2020,ahmed2021quantum,zuo2022optical}, which is especially important for large systems where full characterization is computationally challenging. On the other hand, Hamiltonian learning is an essential technique for estimating Hamiltonians~\cite{Qi2019determininglocal,PhysRevB.100.134201,chen2012ground,PhysRevLett.122.020504,Anshu2021,PhysRevA.105.012615,PhysRevLett.122.150606,Cao2020,Hou2020}, which is crucial for quantum computing and simulation, as it directly affects the control and manipulation of quantum systems. Despite its importance, Hamiltonian learning indeed be challenging due to both data acquisition and computational complexity. 
	
	Although quantum state tomography and Hamiltonian learning have made significant progress separately, a single approach that merges the advantages of both techniques has not been developed yet. Creating a unified method would be a useful addition to quantum information science, as it could allow for more effective and precise characterization and control of complex quantum systems.
	
	Recent breakthroughs in machine learning and natural language processing (NLP), particularly the advent of transformer architectures~\cite{vaswani2017attention}, have profoundly influenced scientific research. The transformer architecture boasts a highly modular design, effortlessly accommodating larger datasets and tackling increasingly intricate tasks~\cite{devlin2018bert,radford2018improving,radford2019language,brown2020language,ouyang2022training}. These sophisticated advancements have been successfully integrated into quantum information studies~\cite{cha2021attention,wang2022predicting,PhysRevB.107.075147,zhong2022quantum}, facilitating substantial progress in comprehending quantum systems.
	
	In this study, we introduce a novel approach that utilizes language translation method to effectively address both quantum state tomography and Hamiltonian learning, uniting these two techniques in a unified model. The attention mechanism within the transformer model enables us to establish a language-translation-like strategy for mapping Hamiltonian parameters to quantum states. We apply our methodology to an extensive spectrum of quantum systems, ranging from 2-qubit cases to 2D antiferromagnetic Heisenberg model, and demonstrate the versatility of our approach by employing various QST methods. A notable example is the classical shadow technique~\cite{aaronson2018shadow,akhtar2022scalable,bertoni2022shallow,acharya2021informationally,hu2021classical,huang2020predicting,huang2021efficient,PhysRevLett.129.220502}, which reduces the computational resources needed for QST, making it more practical for larger systems. 
	
	A notable strength of our approach lies in its capacity to combine QST and Hamiltonian learning tasks without requiring changes to the underlying transformer model's architecture or parameters. This is the first advantage. The second advantage is that the model can effectively learn the complex relationships between quantum states and Hamiltonians, provided that the training data is carefully selected and prepared. Lastly, the third advantage is the simplified data acquisition process, as obtaining training data only necessitates a unidirectional generation process starting from state tomography. These three advantages contribute to the effectiveness and generalizability of our proposed method, allowing it to be applied across a wide range of quantum systems.
	
	Our findings reveal that our approach accurately predicts not only the ground-state measurements of these systems based on the Hamiltonian parameters, but also the Hamiltonian parameters of the systems based on the observed measurements. The scalability and few-shot learning capabilities of our approach highlight the potential reduction of resources needed for characterizing and optimizing quantum systems, underscoring the method's potential for further quantum system research and the advancement of quantum technologies.
	
	The structure of this paper is as follows: In Sec.~\ref{sec:method}, we provide a concise overview of quantum state tomography and Hamiltonian learning, as well as an in-depth description of the proposed method and its implementation. Sec.~\ref{sec:num} presents the numerical results and analysis, focusing on a 2-qubit toy model (Sec.~\ref{sec:2qu}) and a 2D anti-ferromagnetic random Heisenberg model (Sec.~\ref{sec:2d}). The scalable few-shot learning approach for large-scale 2D anti-ferromagnetic random Heisenberg models is discussed in Sec.~\ref{sec:scal}. Finally, Sec.~\ref{sec:con} offers concluding remarks and explores potential avenues for future research in this domain.

	\section{Method}
	\label{sec:method}
	
	\subsection{Quantum State Tomography and Hamiltonian Learning}
	
	In this section, we provide a preliminary overview of the problems under investigation within the domains of quantum state tomography and Hamiltonian learning. Throughout this work, we consider the $k$-local Hamiltonian, an $n$-qubit Hamiltonian that can be expressed as a sum of terms, with each term non-trivially affecting a maximum of $k$ qubits. The Hamiltonian acts on the Hilbert space $\mathcal{H} = \bigotimes_{i=1}^n \mathbb{C}^2$ of an $n$-qubit quantum system. Formally, the Hamiltonian is expressed as follows:
	
	\begin{equation}
		H(\vec{x}) = \sum_{i=1}^m x_i H_i,
	\end{equation}
	
	where each $H_i$ denotes an operator acting non-trivially on no more than $k$ qubits, and $x_i$ represents the parameter of the local terms. In this work, we explore the $k$-local Hamiltonian with $k=2$, which implies that the number of local terms, $m$, is at most of the order of a polynomial in the number of qubits, i.e., $m = \text{poly}(n)$.
	
	In our investigation, we explore the quantum state of the system, specifically concentrating on its ground state.  The state is represented by the density operators $\rho(\vec{x})$. These operators are related to the Hamiltonian $H(\vec{x})$ through the Boltzmann factor and the inverse temperature $\beta = 1/k_B T$, where $k_B$ denotes the Boltzmann constant and $T$ signifies the temperature:

	\begin{equation}
		\rho(\vec{x}) = \frac{e^{-\beta H(\vec{x})}}{Z(\vec{x})}.
	\end{equation}
	Within this framework, $Z(\vec{x})$ represents the partition function, which encompasses the summation of the Boltzmann factors for all states in the ensemble: $Z(\vec{x}) = \mathrm{Tr}[e^{-\beta H(\vec{x})}]$. For the ground state, as $T \to 0$, we have $\beta \to \infty$. 
	
	We consider physical systems composed of $n$ qubits and construct our measurements originating from an $\mathbb{m}$ outcome single-qubit POVM $\mathcal{M}=\{M^{(b)}\}_b$, defined by positive semi-definite operators $M(b)\geq 0$, each uniquely labeled by a measurement outcome $b=0,1, \ldots, \mathbb{m}-1$. These satisfy the normalization requirement $\sum_b M^{(b)}=\mathbb{1}$. The $n$-qubit measurement is characterized by the tensor product of the single-qubit POVM elements $\mathcal{M}=\{M^{\left(b_1\right)} \otimes M^{\left(b_2\right)} \otimes \ldots M^{\left(b_n\right)}\}$. Due to Born's rule, the probability distribution $P(\vec{b}|\vec{x})$ over measurement outcomes $\vec{b}={b_1,b_2, \ldots, b_n}$ on a quantum state $\rho(\vec{x})$, with $P(\vec{b}|\vec{x}) \geq 0$ and $\sum_{\vec{b}} P(\vec{b}|\vec{x})=1$, is given by the linear expression $P(\vec{b}|\vec{x})=\operatorname{Tr}\left[M^{(\vec{b})} \rho(\vec{x})\right]$. This relation can be formally inverted (refer to Appendix), and we employ the method to recover the density matrix of the ground state to obtain the desired measurement outcomes throughout the entire process. By implementing this measurement approach and recovery technique, we are able to acquire the desired measurement outcomes that uniquely determine the ground state of the 2-local Hamiltonian throughout the entire procedure.
	
	In this research endeavor, our focus is on the Pauli-6 POVM, comprising $n$-qubit tensor products of projections onto the eigenspaces of the Pauli observables\cite{carrasquilla2019reconstructing}. The Pauli-6 POVM corresponds to performing measurements in one of the three Pauli bases, uniformly chosen at random. The acquisition of training data requires a unidirectional generation process originating from state tomography, as represented by the transformation $H(\vec{x})\rightarrow P(\vec{b}|\vec{x})$ (see Fig.\ref{fig:translation model}). This streamlined approach not only simplifies the data acquisition process but also confers a significant advantage in the simultaneous training of both QST and Hamiltonian learning, thereby enhancing the efficiency and effectiveness of these interconnected methodologies in elucidating quantum states and estimating Hamiltonian parameters.
	
	Within the domain of quantum state tomography, the primary objective is to estimate a family of states, denoted as $\rho(\vec{x})$, utilizing a dataset $\mathcal{D}=\left(\vec{b}^{(1)}, \vec{x}^{(1)}\right), \ldots,\left(\vec{b}^{(N_s)}, \vec{x}^{(N_s)}\right)$ comprising $N_s$ samples. A generative model $p_\theta$, parameterized by a neural network, is trained to optimize the likelihood of observed measurements. The joint distribution is decomposed into conditional distributions via an autoregressive approach:
	
	\begin{equation}
		p_{\theta}\left(b_1, \ldots, b_n \mid \vec{x}\right)=\prod_{i=1}^n p_\theta\left(b_i \mid b_{i-1}, \ldots, b_1, \vec{x}\right).
	\end{equation}
	
	The estimation of the ground state for a Hamiltonian $H(\vec{x})$ is predicated upon measurement outcomes $\vec{b}$. State tomography is effectuated through two distinct methodologies: direct POVM measurement and classical shadow, with supplementary information provided in the Appendix.
	
	In the realm of Hamiltonian learning, the emphasis is placed upon the ground states $\rho(\vec{x})$ of quantum systems that exhibit a particular structure. The Hamiltonian parameters are represented by a vector $\vec{x}=\left(x_1, \ldots, x_n\right)$, akin to the previously mentioned measurement outcomes. The dataset, $\mathcal{D}=\left(P(\vec{b}^{(1)}|\vec{x}^{(1)}), \vec{x}^{(1)}\right), \ldots,\left(P(\vec{b}^{(N_s)}| \vec{x}^{(N_s)}), \vec{x}^{(N_s)}\right)$, comprises the probability distribution of measurement outcomes $P(\vec{b})$. The objective is to deduce Hamiltonian parameters $\vec{x}$ corresponding to a given ground state $\rho(\vec{x})$ based on the observations $\vec{b}$:
	
	\begin{equation}
		p_{\theta}\left(x_1, \ldots, x_n \mid P(\vec{b}|\vec{x})\right)=\prod_{i=1}^n p_\theta\left(x_i \mid x_{i-1}, \ldots, x_1, P(\vec{b}|\vec{x})\right).
	\end{equation}
	
	\subsection{Translator Model}
		\begin{figure*}
		\includegraphics[width=\textwidth]{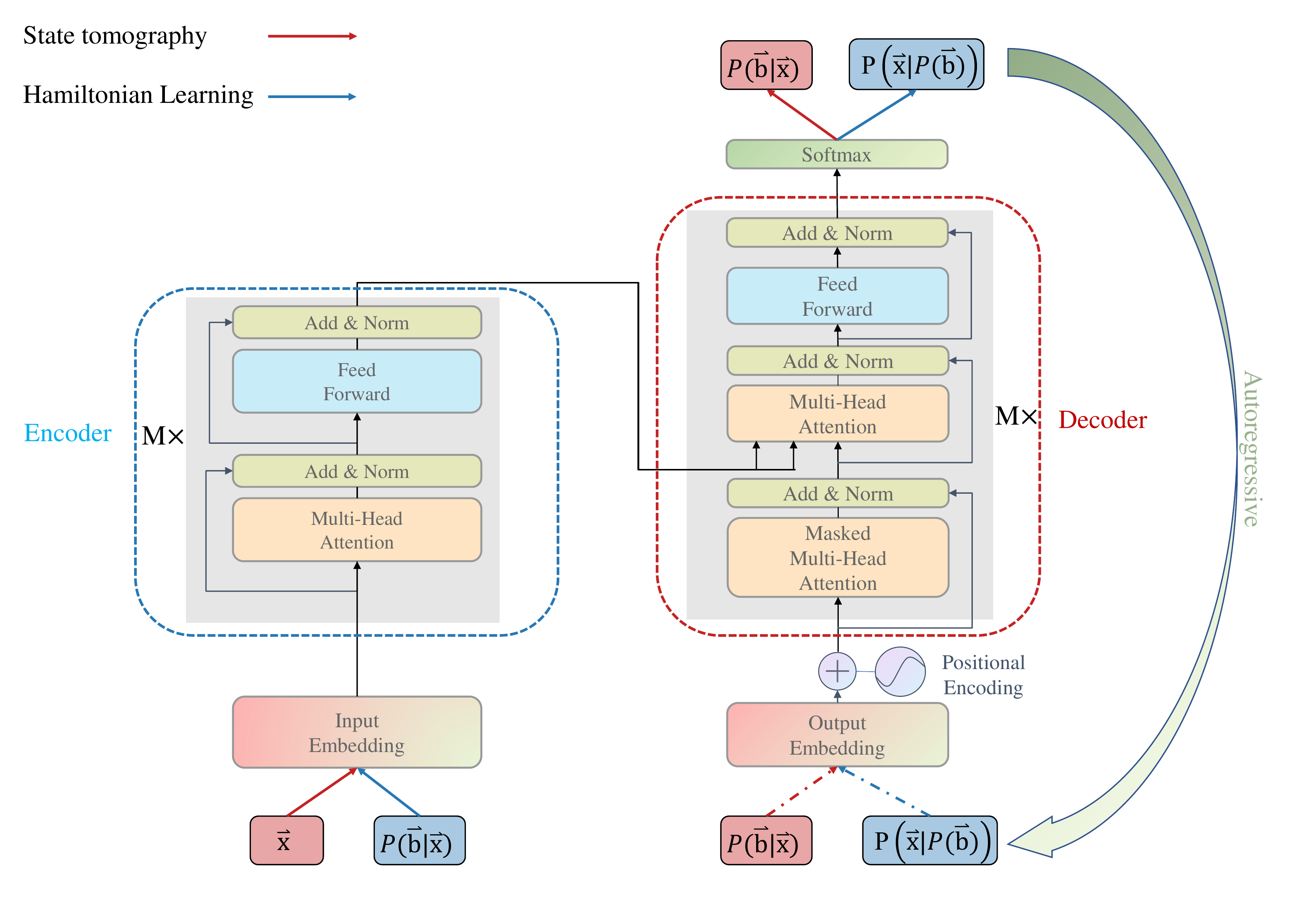}
		\caption{Bidirectional Refined Transformer Architecture: The model consists of an encoder and decoder, with the encoder designed for processing continuous Hamiltonian parameters $\vec{x}$ or local measurement probability distributions $P(\vec{b}|\vec{x})$, replacing the conventional word embedding layer with a multilayer neural network. The decoder follows the original transformer design, handling discrete inputs like discretized Hamiltonian parameters $\vec{x}$ or vocabulary tokens for quantum state tomography outcomes $\vec{b}$. Dashed arrows represent sampling from a given distribution. Auto-regression is used for output comparison, with prior observations informing predictions. Notably, only input and output data modifications (red and blue arrows) are needed, preserving the core model architecture.}
		\label{fig:Trans}
	\end{figure*}
	
	In this study, we introduce a sophisticated model based on the Transformer architecture, ingeniously engineered to establish a bidirectional relationship between Hamiltonian parameters and the measurement outcomes associated with the system's ground state. As depicted in Fig.~\ref{fig:Trans}, the proposed model encompasses an encoder and a decoder, wherein the encoder skillfully transmutes continuous variable inputs into concise, fixed-length representations that encapsulate the input correlations. Moreover, the decoder generates the target sequence output, which may incorporate discrete Hamiltonian parameters or measurement expressions in hitherto undefined languages.
	
	To accomplish the intended outcome, our model employs an embedding neural network that transforms each input into a vector representation, encapsulating the essence of individual elements. The model then leverages self-attention to concentrate on various portions of the encoder input while generating the final output, thereby capturing the most pertinent information from the inputs when producing the target measurements or parameters.

	In our approach, we employ a multilayer neural network as the embedding layer for those two tasks, transforming continuous Hamiltonian parameters $\vec{x}$ or probability distribution $P(\vec{b}|\vec{x})$ of local measurements into learned vector representations. Since the POVM sequence $\vec{b}$ is discrete and amenable to linguistic processing, and we discretize the Hamiltonian parameter $\vec{x}$, we utilize the word embedding technique, widely used in natural language processing (NLP), to map each label to its corresponding learned vector representation. In our approach, we encode distinct local measurement outcomes into discrete tokens in a vocabulary list $\mathcal{V}$, with $|\mathcal{V}| = N_m + 3$,
	\begin{equation}
		\mathcal{V} \equiv\{\text { sos, eos, pad }, 3, \ldots, N_m+3\}.
		\label{eq:vocab}
	\end{equation}
	This list is then used to encode any measurement outcome or parameter of an $n$-qubit quantum system into a word sequence. Ultimately, the model generates the probability distribution $P(\vec{b}|\vec{x})$ for the ground state under varying parameters of the given Hamiltonian $H(\vec{x})$ or generates the probability distribution $P(\vec{x}|P(\vec{b}))$ for the Hamiltonian parameters of the given probability observation $P(\vec{b}|\vec{x})$ along with its ground state.
	
	It is important to note that in both tasks there is no need for any modifications to the underlying architecture or parameters of the model (as illustrated in Fig.~\ref{fig:Trans}). The only requirement for the successful application of this approach is the careful selection and preparation of the training data, which plays a crucial role in enabling the model to effectively learn the intricate relationships between quantum states and Hamiltonian.
	
	Throughout the training process, we implement teacher forcing to train our translation model. Teacher forcing is a prevalent training technique for neural machine translation that utilizes the actual output rather than the predicted output from the previous timestamp as inputs during training. This approach expedites the training process. In our problem, we input the training data into the encoder and the training label into the decoder. The output is compared through auto-regression, which refers to a time series model that employs observations from previous time steps as input to a regression equation to predict the value at the next time step.
	
	The training objective involves minimizing the average negative log-likelihood loss across the training data, as follows:
	\begin{equation}
		\min_{\theta} \mathcal{L}(\theta):=\frac{1}{N_s} \sum_{\mathcal{D}}-\log p_{\theta}.
	\end{equation}

	\section{Numerical Results}\label{sec:num}
	We now apply our algorithm to study some quantum state tomography and Hamiltonian estimation problems. 
	
	\subsection{2-qubit Toy Model}\label{sec:2qu}
	
	In this section, we present a simplistic model to elucidate the dynamics of a two-qubit Hamiltonian. This model encompasses a singular scenario with only one parameter, and the Hamiltonian is expressed by Eq.~\ref{eq:1}, incorporating both the Pauli-X and Pauli-Z operators.
	
	\begin{equation}
		H(\mathbf{\theta})=\cos(\theta) X_1 X_2 +\sin(\theta) Z_1 I_2.
		\label{eq:1}
	\end{equation}
	
\begin{figure*}
		\begin{subfigure}[b]{0.45\textwidth}
			\includegraphics[width=\textwidth]{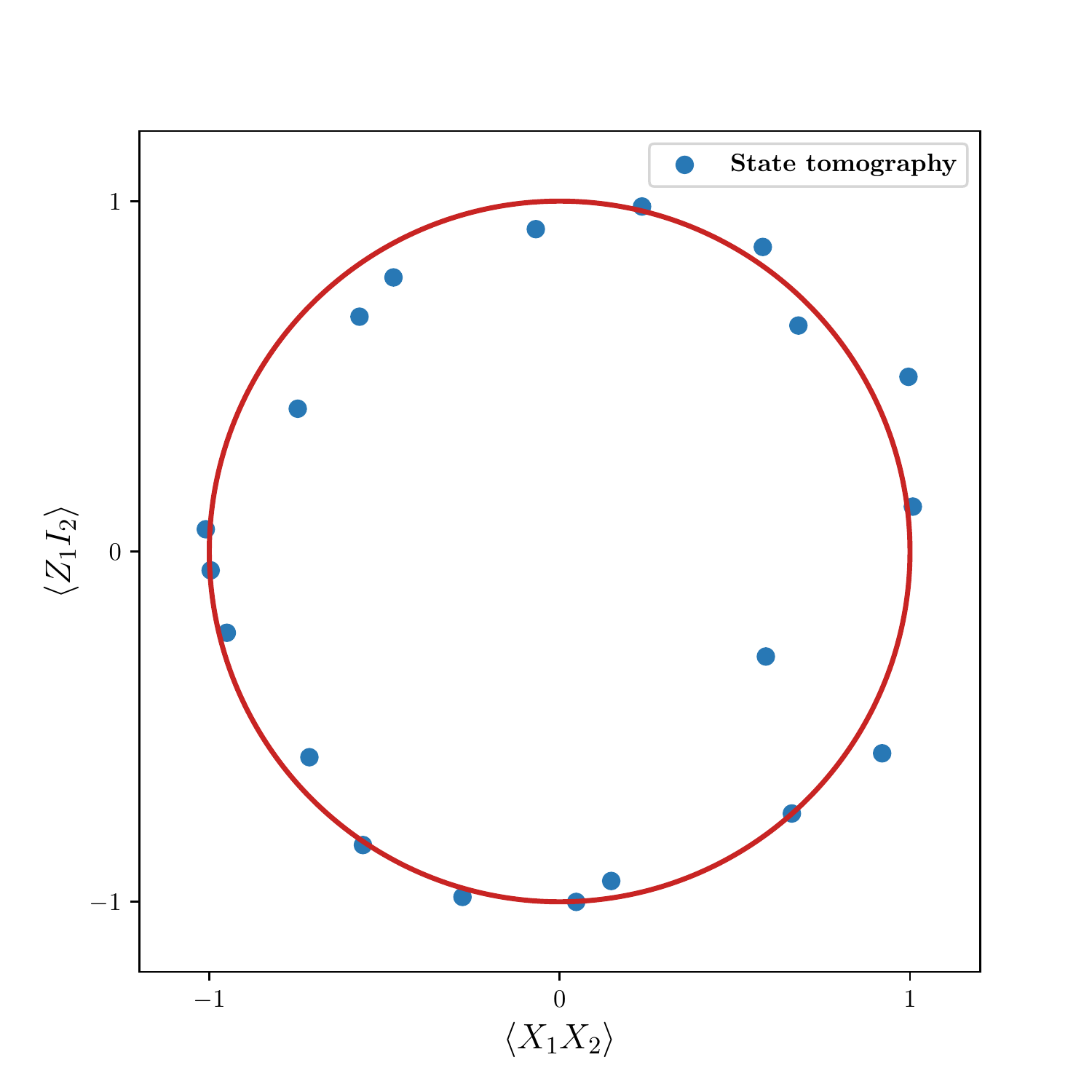}
			\caption{State tomography result}
			\label{fig:tomo_circle}
		\end{subfigure}
		\begin{subfigure}[b]{0.45\textwidth}
			\includegraphics[width=\textwidth]{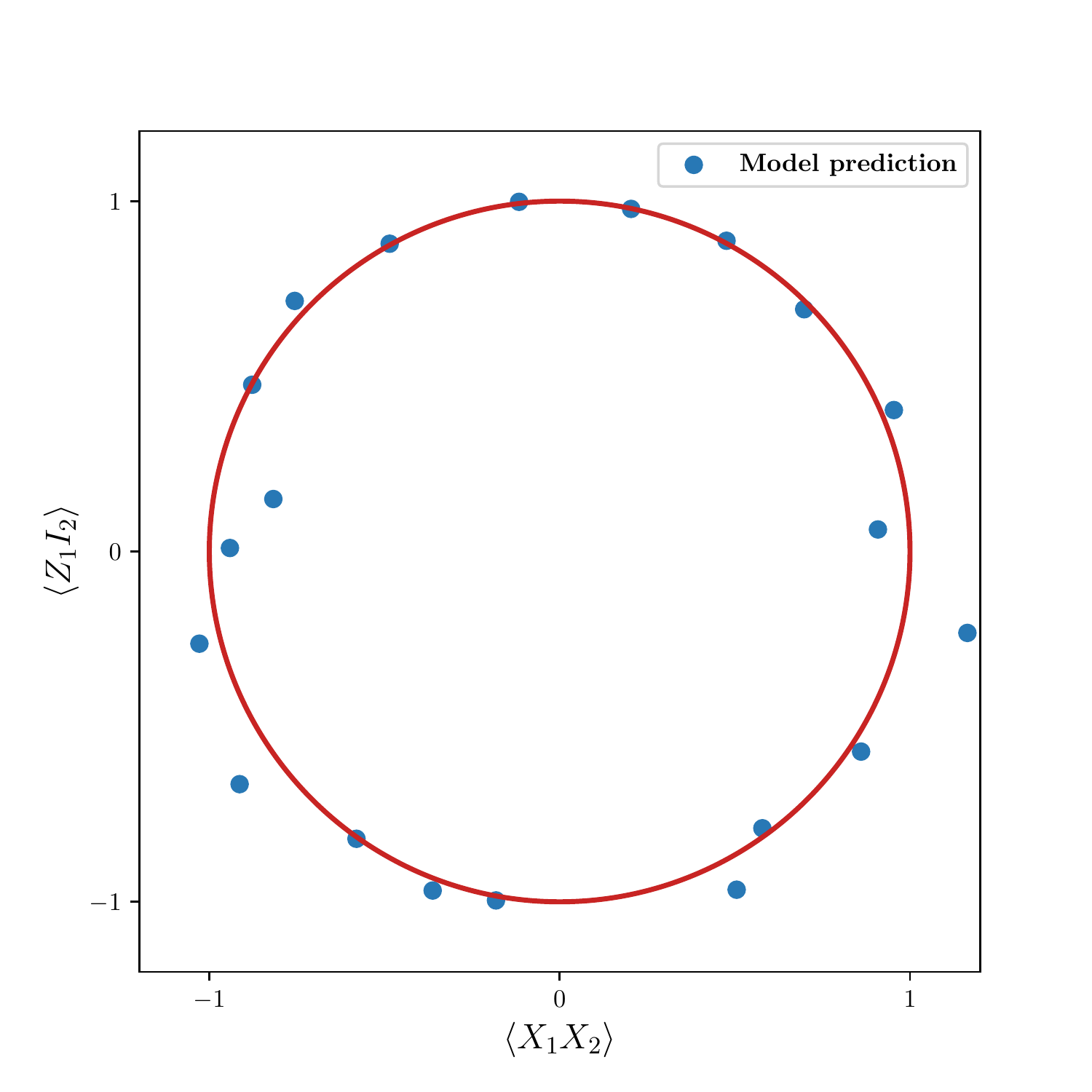}
			\caption{Model prediction}
			\label{fig:trans_circle}
		\end{subfigure}
		\caption{Comparison of state tomography (a) with the predicted outcomes of our model (b), with the expectation value of the operator $X_1 X_2$ on the horizontal axis and the expectation value of the operator $Z_1 I_2$ on the vertical axis. The red circles represent the theoretical expectation values, while the blue dots signify the expectation values of the reconstructed quantum states on the operators}
		\label{fig:1}
	\end{figure*}

	The parameter $\theta$ governs the interplay between the two terms, effectively dictating the system's behavior. Notably, when projected onto the basis composed of Pauli operators $X_1X_2$ and $Z_1I_2$, with the subscript denoting the qubit number, the ground state adopts a circular configuration, as depicted in Fig.~\ref{fig:1}. Our primary objective is to investigate the extent to which the POVM measurements, generated by our model-based learning approach, can be honed to accurately represent the genuine ground state when projected onto the specified basis. This is to be accomplished while accommodating arbitrary operator coordinate systems and adjustments to the system parameters.
	
	We extract 100 uniformly distributed data points from the interval $\theta \in [0, 2\pi]$ to obtain the learning data. The training and testing data sets comprise 80 and 20 data points, respectively. For each $\theta$, we collect 1000 ground state measurements, resulting in an $N_s = 100000$ data set of randomized Pauli measurements. Our experimental results, showcased in Fig.~\ref{fig:1}, indicate that the trained model achieves exceptional accuracy in estimating the ground state, even without knowledge of the Hamiltonian parameters. Furthermore, by reconstructing the measurements into quantum states and projecting them onto the designated space, we establish that the relationship can be approximated as a circle.

\begin{figure}
		\centering
		\includegraphics[width=0.5\textwidth]{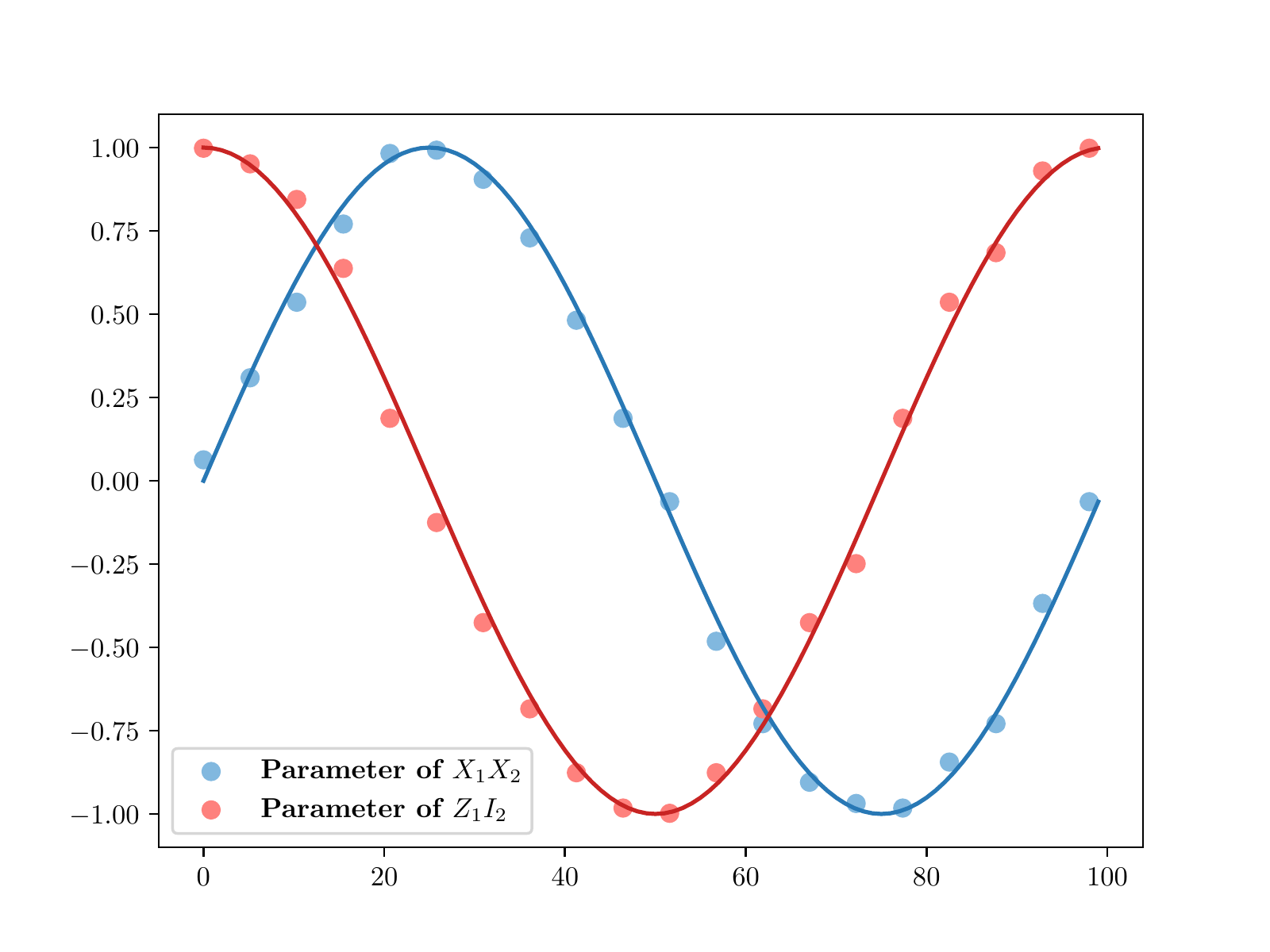}
		\caption{Prediction results of our model on the Hamiltonian parameter. The vertical axis is the value and the horizontal axis is the index number of the total data set. We sampled 20 of the 100 total data sets as test data.
			The solid lines are the real Hamiltonian parameters and the dots are the predictions of our model.}
		\label{fig:reverse_circle}
	\end{figure}

	Following this, we reformulate the training data to facilitate Hamiltonian learning, with the aim of determining whether the algorithm can generate the appropriate Hamiltonian parameters when provided with the ground-state distribution comprising varying Hamiltonian parameters. To achieve this, we utilize a statistical distribution of 1000 samples as input, while the output comprises discrete Hamiltonian parameters corresponding to local terms $X_1 X_2$ and $Z_1 I_2$, resulting in a dataset mirroring the QST task with $N_s=100$. The Hamiltonian parameters are characterized by a vocabulary $N_m$ of size 50. Fig.~\ref{fig:reverse_circle} confirms that our algorithm yields a negligible deviation between the output and the true Hamiltonian parameters for a given distribution of unknown quantum states.
	
	\subsection{2D anti-ferromagnetic random Heisenberg model}\label{sec:2d}
	In the second example explored, we investigate the two-dimensional (2D) antiferromagnetic random Heisenberg model, in which qubits—specifically, spin-$1/2$ particles—are arranged on a square lattice. We primarily focus on the ground state of the Hamiltonian:
	\begin{equation}
		H(\mathbf{x})=\sum_{\langle i j\rangle} \mathbf{x}_{i j}\left(X_i X_j+Y_i Y_j+Z_i Z_j\right),
		\label{eq:heisenberg}
	\end{equation}
	where $\langle ij\rangle$ denotes nearest-neighbor interactions, and the summation encompasses all possible pairs on the lattice. For each pair $\langle ij\rangle$, the corresponding interaction strength $x_{i j}$ is uniformly sampled from the interval $[0,2]$. The Hamiltonian in Eq.~\ref{eq:heisenberg} can be represented by a weighted undirected graph without any loss of information. Each qubit appears as a node, and the coupling strength between two sites corresponds to a weighted edge in the graph. Denoting the adjacency matrix of the graph as $g(\vec{x})$, we designate $\rho(g(\vec{x}))$ for the ground states of the Hamiltonian corresponding to the grid lattice defined by the graph $g(\vec{x})$.
	
	\begin{figure*}
		\begin{subfigure}[b]{0.3\textwidth}
			\raisebox{0.5cm}{\includegraphics[width=0.9\textwidth]{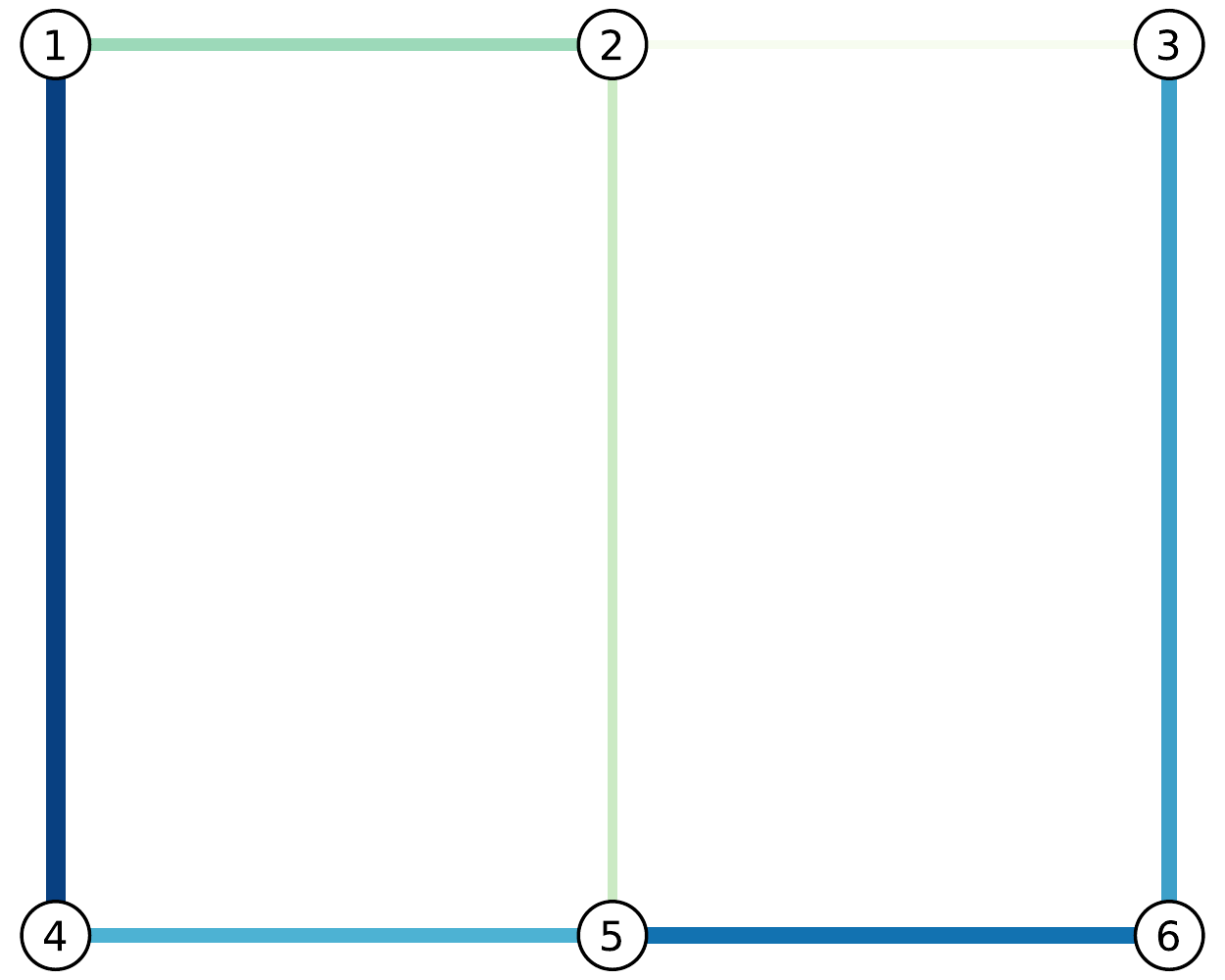}}
			\caption{Coupling graph}
			\label{fig:J}
		\end{subfigure}
		\begin{subfigure}[b]{0.6\textwidth}
			\includegraphics[width=\textwidth]{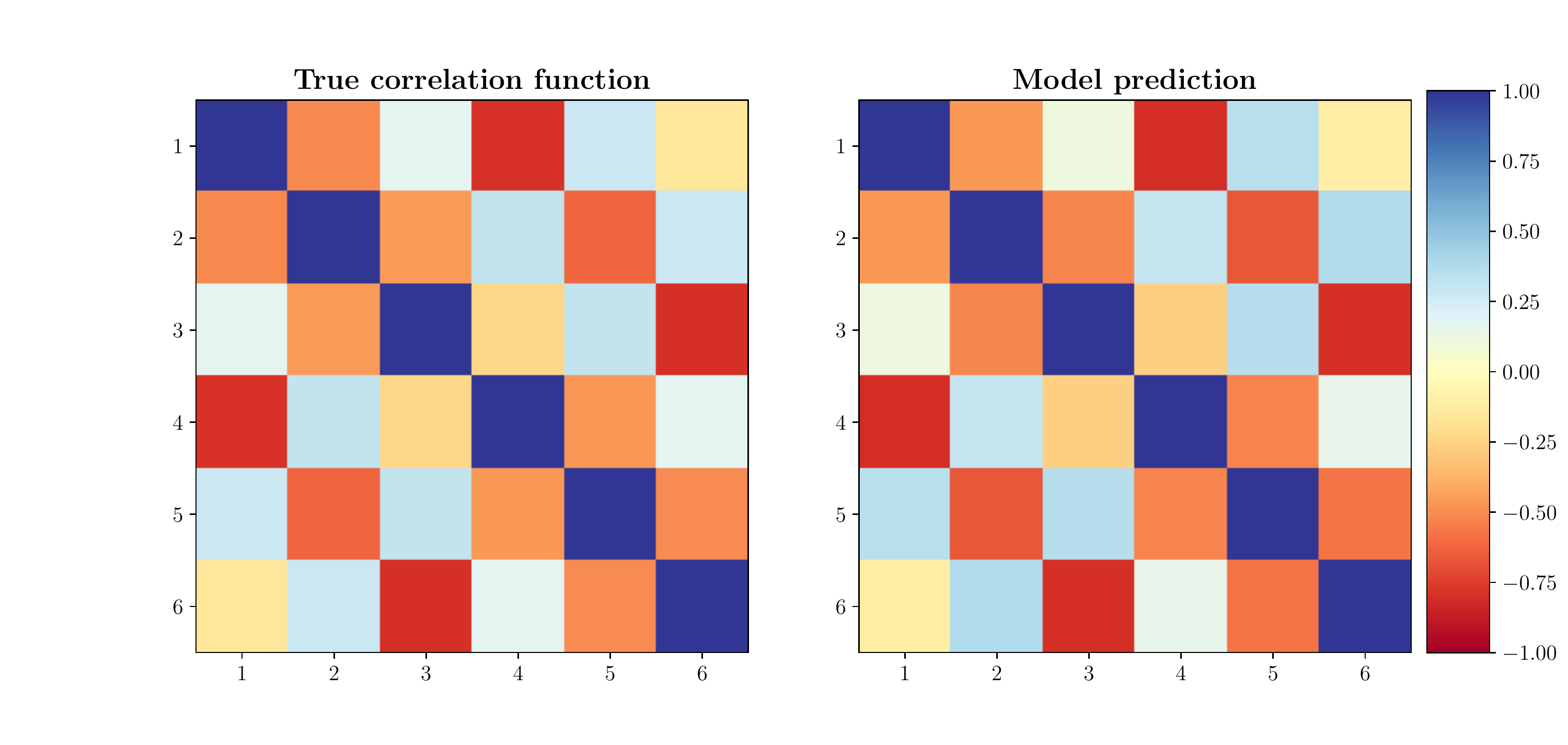}
			\caption{Two-point correlation functions}
			\label{fig:trans_corre}
		\end{subfigure}
		\caption{Predicting correlation functions of ground states of the 2D random antiferromagnetic Heisenberg model. (a) A stochastic coupling graph from the test set defining the 2D random Heisenberg model Eq.\ref{eq:heisenberg} and utilized for translation in our model. The thickness and color of the edges in the graph indicate the strength of the interaction, with thicker and darker edges representing higher interaction strengths. (b) A comparison between the authentic and predicted two-point correlation functions Eq.\ref{eq:corre} for a ground state from the test set, encoded by our translator model given the coupling graph.}
		\label{fig:2}
	\end{figure*}
	
	Taking into account the exponential relationship between accuracy and the number of measurements for multi-qubit systems, as well as the significance of physical observable behavior in quantum many-body systems, we utilize the classical shadow methodology in this study to observe and retrieve the physical observables of interest. In the context of the quantum state tomography task, we assess the similarity between the predicted quantum state, obtained from the classical shadow of the ground state, and the true state by scrutinizing the expectation value of the two-point correlation function, denoted as $\langle C_{ij} \rangle$.
	\begin{equation}
		C_{ij}=\frac{1}{3}(X_iX_j+Y_iY_j+Z_iZ_j).
		\label{eq:corre}
	\end{equation}

	As a point of comparison, we utilize the classical machine learning kernel approach to predict the correlation functions of the test set. To gauge the difference between the methods, we select the root mean square error (RMSE) for comparison:
	\begin{equation}
		\text{RMSE}=\sqrt{\frac{\sum_{i,j}\left(\langle C^\text { Predicted }_{ij} \rangle-\langle C^\text { Actual }_{ij}\rangle \right)^2}{n^2}}.
		\label{eq:rmse}
	\end{equation}
	
	The dataset utilized for training and evaluation is derived from classical simulations. We generate 100 random Hamiltonians by sampling coupling constants uniformly at random, $x_{ij} \stackrel{\mathrm{iid}}{\sim} \mathcal{U}[0, 2]$, and ascertain the ground state through exact diagonalization. For each randomly sampled Hamiltonian, we gather 1000 ground state measurements, yielding a dataset consisting of 100000 randomized Pauli measurements. In a manner akin to the previous section, we train our model on 80 Hamiltonians and allocate the remaining 20 Hamiltonians to the test set in order to assess the generalization capabilities of our models with respect to unfamiliar lattice structures.
	\begin{figure}
		\centering
		\includegraphics[width=0.5\textwidth]{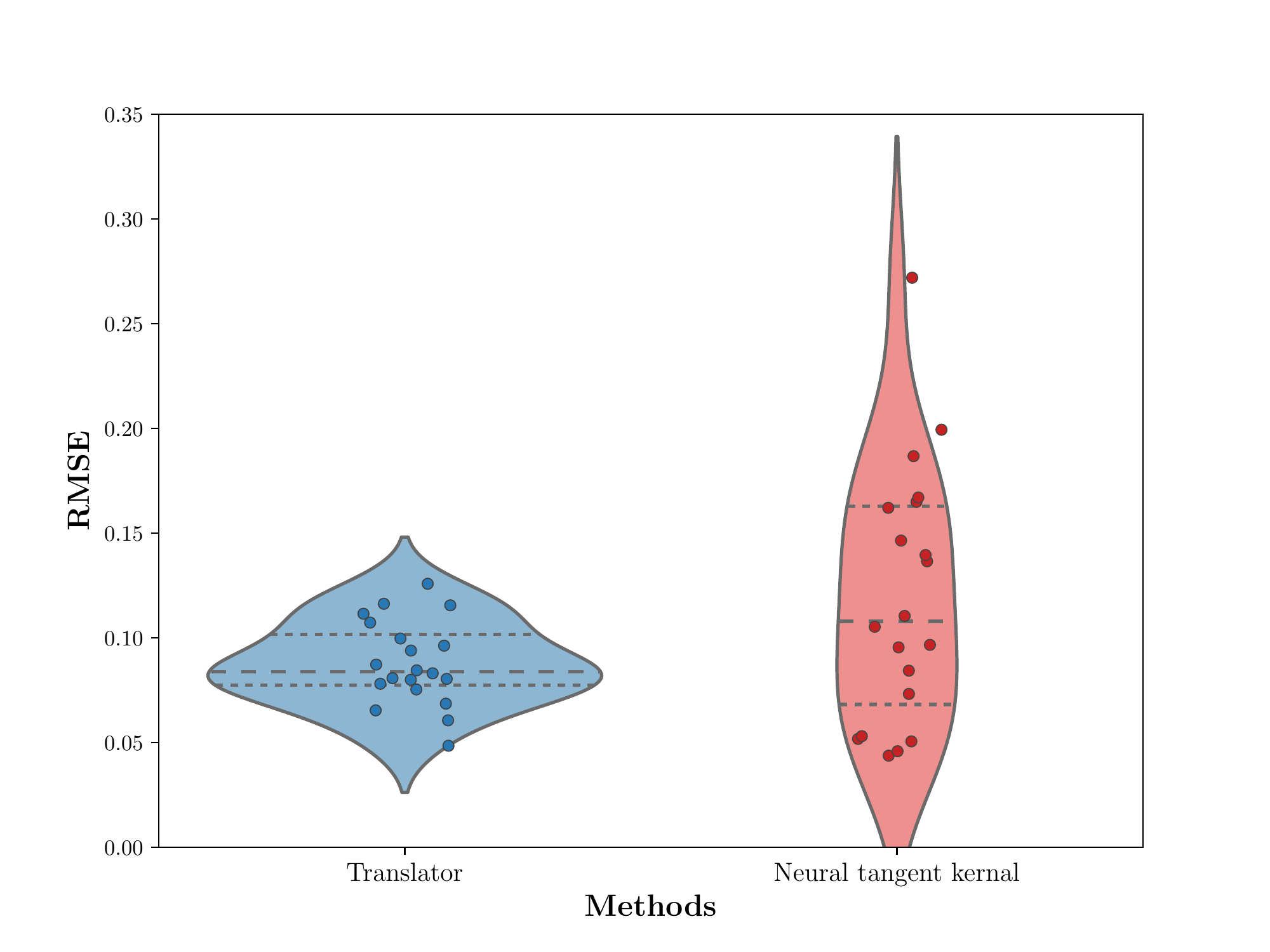}
		\caption{Violin plot with stripplot of RMSE between authentic and estimated correlation
			functions, for our translator model (blue) and Neural Tangent Kernel (red). Each point in the plot signifies the error of correlation predictions from the test set. Three dashed lines represent the quartiles of the data distribution. The middle dashed line is the median ($50\%$), while the other two dashed lines are the lower quartile ($25\%$) and the upper quartile ($75\%$). The outline of the violin plot represents the estimated probability density of the data at different values.}
		\label{fig:RMSE_2_3}
	\end{figure}

	In Fig. \ref{fig:trans_corre}, we display both the genuine and predicted correlation functions corresponding to the Hamiltonian characterized by the random coupling graph depicted in Fig. \ref{fig:J}. These predictions are obtained by conditioning the translation model on the coupling graph and subsequently generating new samples (i.e., POVM outcomes) to reconstruct a classical shadow. Using the obtained classical shadow, we estimate the observables related to the correlation functions by employing Eq.\ref{eq:corre}. In Fig.\ref{fig:RMSE_2_3}, we present the root-mean-square error (RMSE) between the predicted and actual correlation functions for various methods. Each data point in the figure represents the prediction error observed in the test set. We also include the prediction inaccuracies for the Neural Tangent Kernel approach~\cite{Huang2022}, specifically designed to predict the function $\vec{x} \mapsto\left\langle C_{i j}\right\rangle_{\rho(\vec{x})}$ for a particular pair ${i, j}$ (i.e., requiring the training of a separate model for each site pair ${i, j}$). The figure demonstrates that the model can effectively encode a state with a correlation function closely resembling the true correlation function, exhibiting a high level of accuracy.

	To explore the task of Hamiltonian learning, we examine the adjacency matrix generated by the coupling graph corresponding to the Hamiltonian. We employ a comparative analysis using the root-mean-square error (RMSE), as expressed in Eq.~\ref{eq:rmse}. This approach enables us to evaluate the efficacy of our model in predicting Hamiltonians and their properties, emphasizing its potential contributions to the field of quantum many-body systems and providing valuable insights for both theoretical and experimental studies.

\begin{figure*}
		\begin{subfigure}[b]{0.5\textwidth}
			\raisebox{0.4cm}{\includegraphics[width=\textwidth]{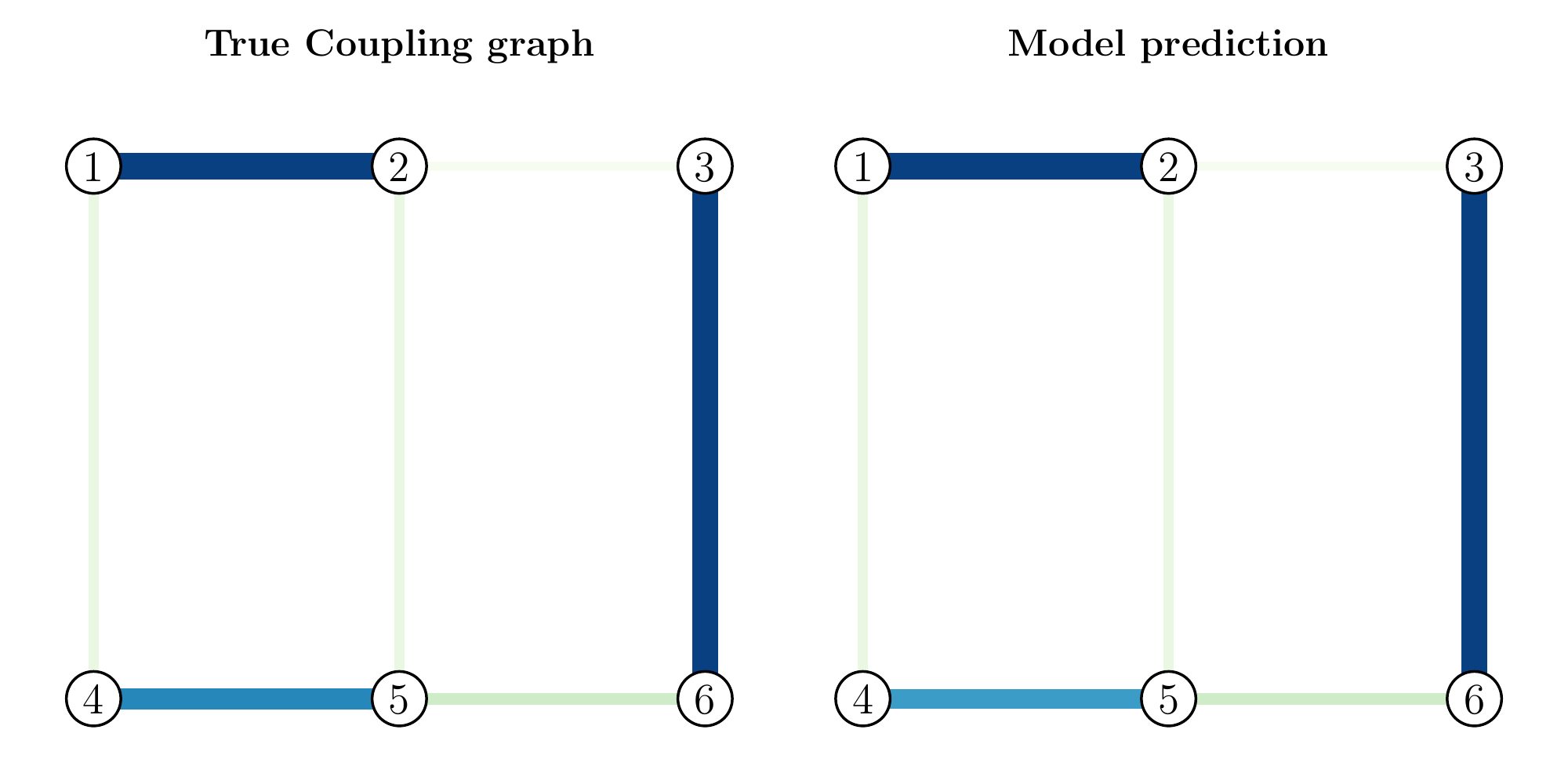}}
			\caption{Coupling graphs}
			\label{fig:rev}
		\end{subfigure}
		\begin{subfigure}[b]{0.45\textwidth}
			\includegraphics[width=\textwidth]{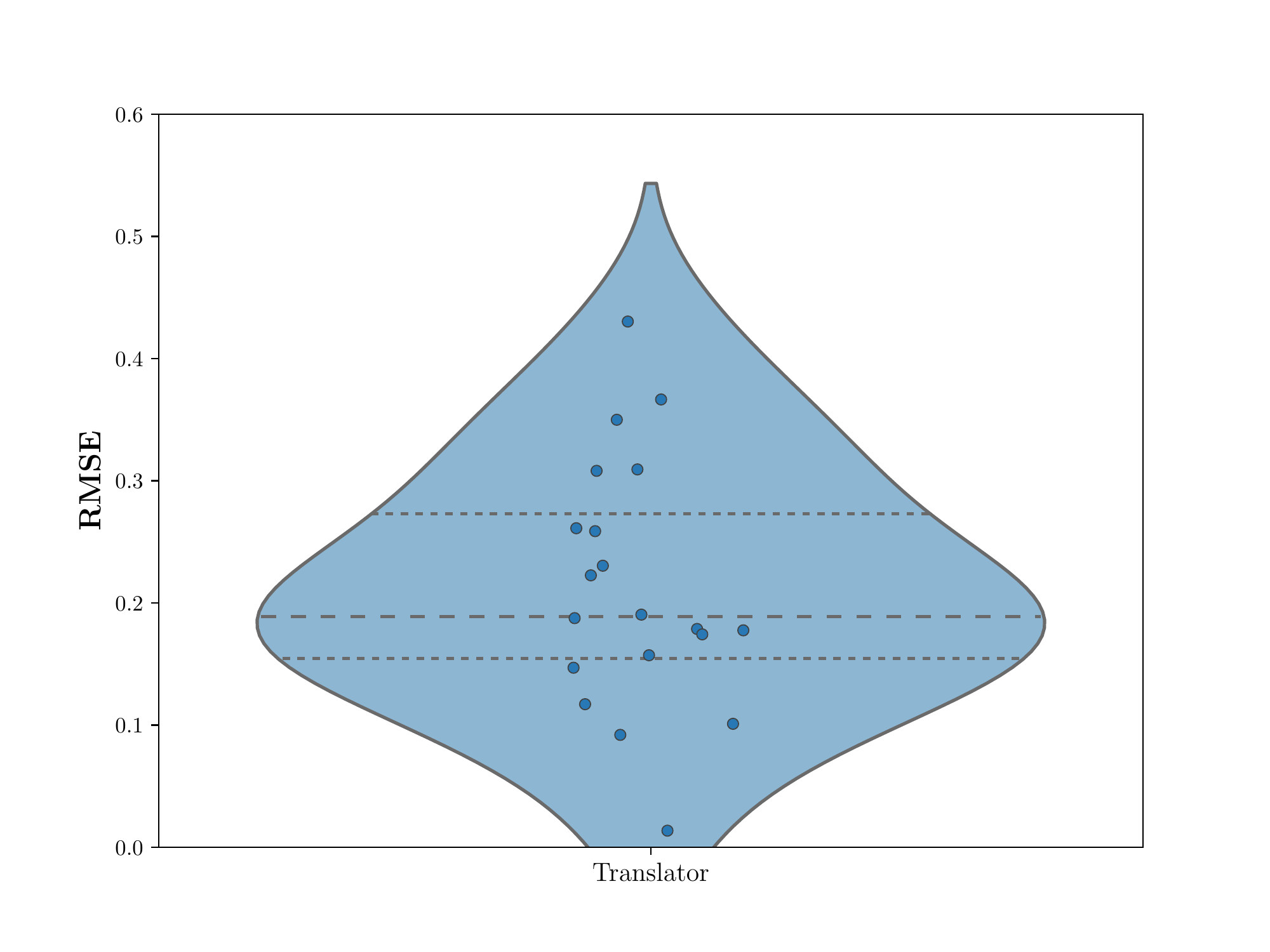}
			\caption{Violin plot with stripplot of RMSE between authentic and estimated adjacency matrices}
			\label{fig:rev_comp}
		\end{subfigure}
		\caption{Forecasting Hamiltonian parameters of the 2D random antiferromagnetic Heisenberg model. (a) A comparison between the authentic and predicted coupling graph for a ground state from the test set, encoded by our translator model given the measurement distribution. The thickness and color of the edges in the graph indicate the strength of the interaction, with thicker and darker edges representing higher interaction strengths. (b) Root Mean Square Error (RMSE) between true and predicted adjacency matrices.}
		\label{fig:3}
	\end{figure*}

	In a practical context, we obtain coupling constants by sampling them in an independent and identically distributed manner, $x_{i j} \stackrel{\mathrm{iid}}{\sim} \mathcal{U}[0, 2]$, wherein each interaction strength is discretized using $N_m = 20$. This results in a parameter space of Hamiltonian parameters on the order of $20^7 \approx 10^9$. Consequently, we collect a total dataset of 100000 Hamiltonians by sampling this quantity, ensuring consistency with the QST task. We assign 99980 samples to the training set, while reserving a mere 20 for the testing set. In Fig.\ref{fig:rev}, we exhibit the authentic and predicted coupling graph for the Hamiltonian in the test set. In Fig.\ref{fig:rev_comp}, we illustrate the RMSE between the predicted and true adjacency matrices from the coupling graph for various methods. Each point in the figure corresponds to the prediction error in the test set. As the majority of techniques in Hamiltonian learning are difficult to compare, given that the training set also functions as the test set in our case, we exclusively provide the error associated with our approach. The figure reveals that the model can effectively encode the coupling graph with a high degree of accuracy, demonstrating its robustness and applicability in predicting the underlying structure of Hamiltonians.
	
	\subsection{Salable Few-Shot Learning} \label{sec:scal}

	In this section, we implement a scalable few-shot learning strategy to predict properties of large-scale models using a limited amount of training data.

	Initially, we take advantage of the easy accessibility of small-scale model data to educate our translation models on these smaller-scale examples. Specifically, we use ground state measurements from the magnitudes of the Hamiltonian with $2\times2$, $2\times3$, and $2\times4$, with 100 ground states each, totaling 300 ground states for learning. Subsequently, we select 20 ground states from $2\times5$ random Hamiltonians for further training and 10 for testing. This process incorporates a limited number of samples with target data, exemplifying the few-shot learning approach.

\begin{figure*}
		\begin{subfigure}[b]{0.35\textwidth}
			\raisebox{0.45cm}{\includegraphics[width=0.8\textwidth]{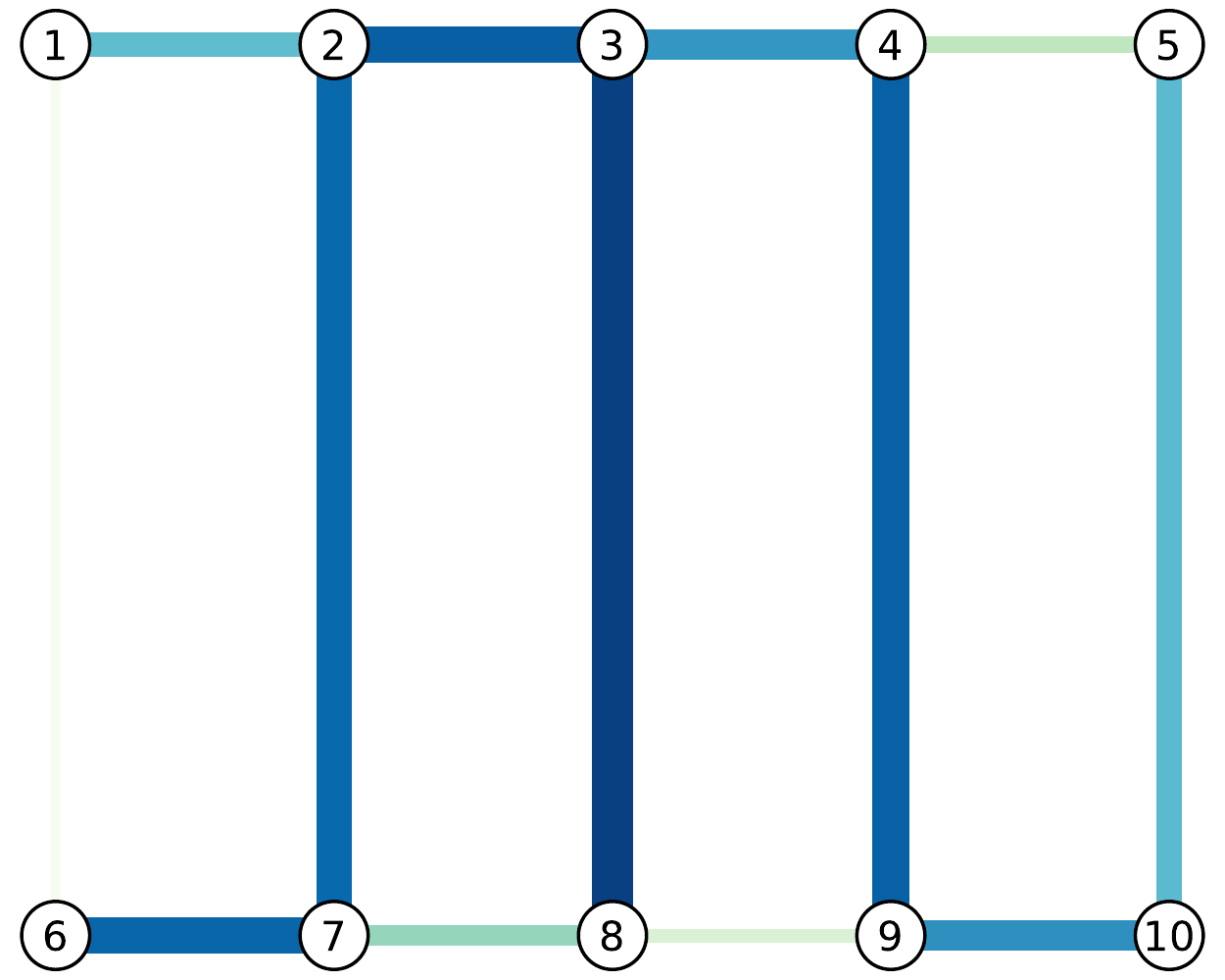}}
			\caption{Coupling graph}
			\label{fig:J_scale}
		\end{subfigure}
		\begin{subfigure}[b]{0.6\textwidth}
			\includegraphics[width=\textwidth]{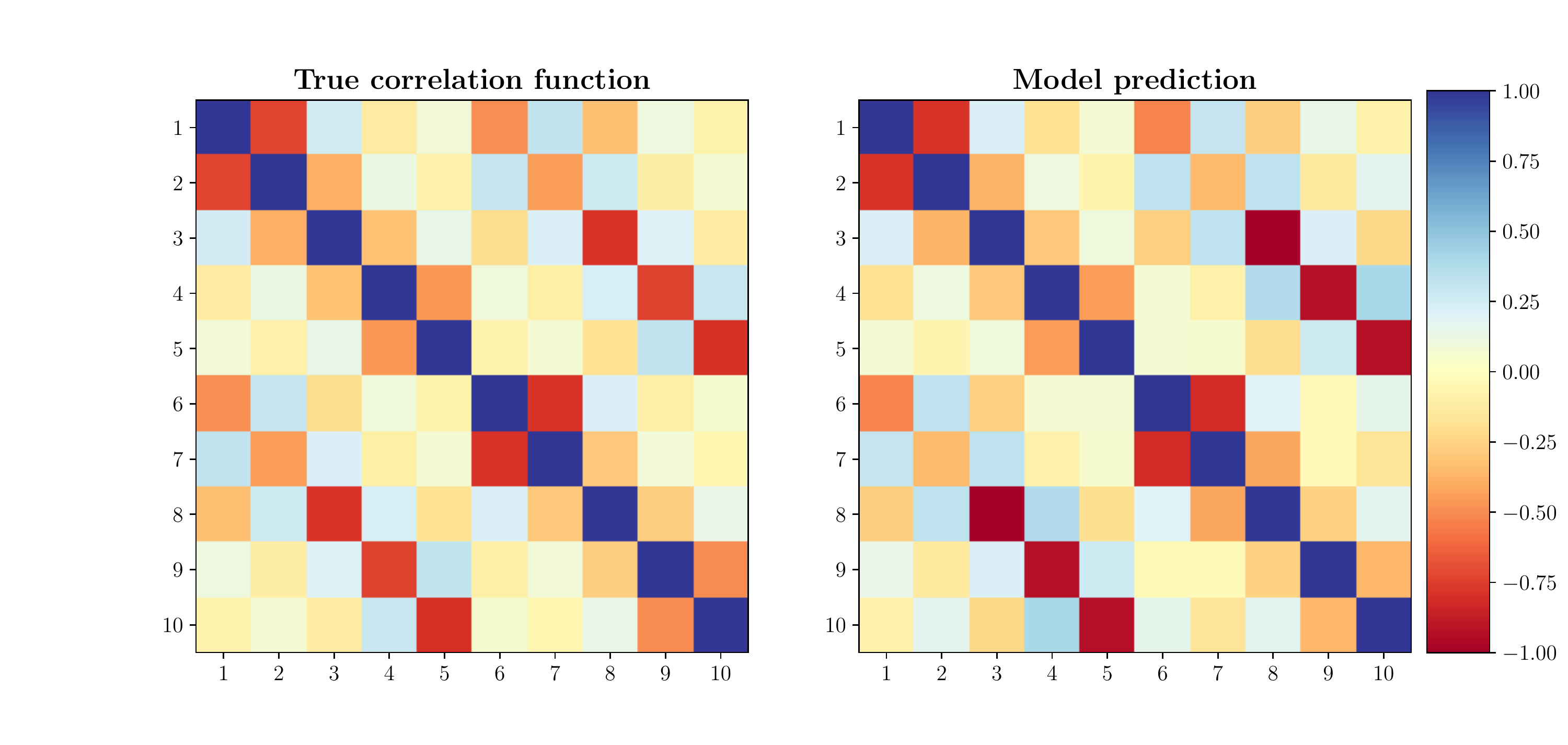}
			\caption{Two-point correlation functions}
			\label{fig:trans_corre_sclae}
		\end{subfigure}
		\caption{Predicting correlation functions of ground states of the 2D random anti-ferromagnetic Heisenberg model. (a) a random coupling graph from the test set that determines the 2D random Heisenberg model~\ref{eq:heisenberg} and is used to translation in our model. The thickness and color of the edges in the graph indicate the strength of the interaction, with thicker and darker edges representing higher interaction strengths. (b) comparison between the true and predicted two-point correlation functions for a ground state from the test set, which is encoded by our translator model given the coupling graph.}
		\label{fig:scale}
	\end{figure*}

\begin{figure}
	\centering
	\includegraphics[width=0.5\textwidth]{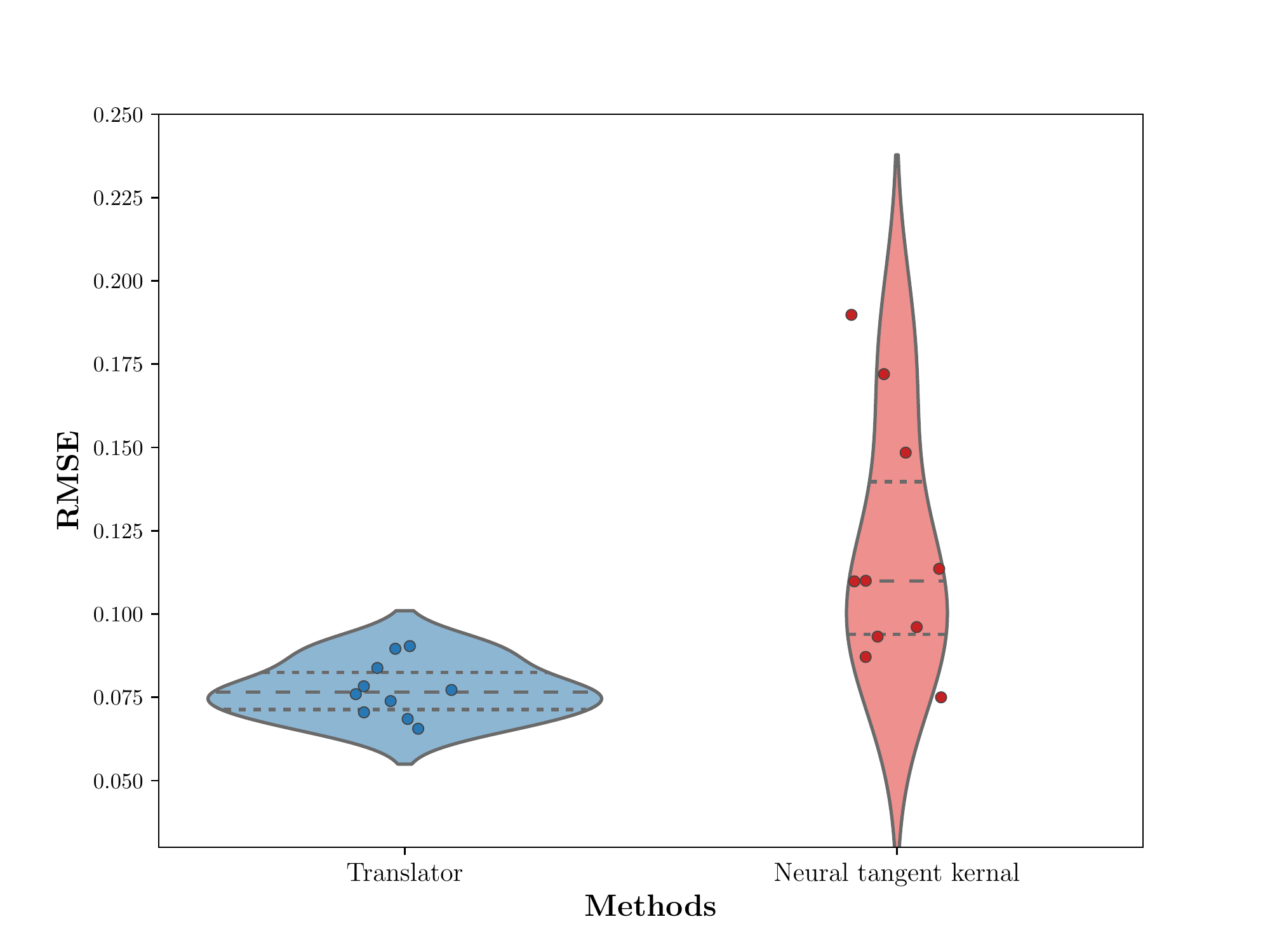}
	\caption{Violin plot with stripplot of RMSE between authentic and estimated correlation
		functions, for our translator model (blue) and Neural Tangent Kernel (red).}
	\label{fig:RMSE_2_5}
	\end{figure}

	As depicted in Fig.\ref{fig:scale} and Fig\ref{fig:RMSE_2_5}, our translation model demonstrates its ability to generate predictions for larger-scale models after being trained on a sparse dataset. This notable capability is achieved through the extrapolation skills developed in the model via its education on the small examples, which equips the model with the ability to generalize to configurations of greater scale and complexity. The effectiveness and versatility of this scheme are evidenced by the model's proficiency in predicting the two-point correlation functions for the unseen Heisenberg model of scale $2\times5$, producing predictions consistent with the ground truths.

	\section{Conclusion}\label{sec:con}
	In this study, we demonstrate a novel approach that utilizes language translation models to effectively tackle quantum state tomography and Hamiltonian learning in a unified framework, an idea not previously investigated. By leveraging the inherent attention mechanism in transformer models, our method unifies QST and Hamiltonian learning tasks without the need for modifications to the underlying model's architecture or parameters. The single requirement is the careful preparation and selection of training data, enabling the model to proficiently decipher the intricate relationships between quantum states and Hamiltonians.
	
	Our approach exhibits the ability to adapt to learning from various techniques while simultaneously reducing the computational resources needed for both quantum state tomography and Hamiltonian learning. The data acquisition process is streamlined, necessitating a unidirectional generation process originating from state tomography. This strategy bypasses the obstacles associated with acquiring data for Hamiltonian learning tasks. The successful application of our method to a diverse range of quantum systems, encompassing elementary 2-qubit scenarios and complex 2D antiferromagnetic Heisenberg models, establishes a promising theoretical basis for pursuing practical quantum advantages with machine learning.
	
	Another significant benefit of our approach resides in its scalability and few-shot learning capabilities, which create opportunities for potentially minimizing the resources required for characterizing and optimizing quantum systems. Additionally, our method furnishes valuable insights into the interplay between Hamiltonian structure and quantum system behavior, an essential aspect for understanding and advancing innovative quantum technologies. As quantum systems grow in size and complexity, the ability to efficiently learn and predict relationships between quantum states and Hamiltonians becomes increasingly vital. In conclusion, our work contributes to the field of quantum information by introducing an innovative, unified, and scalable technique for QST and Hamiltonian learning. This establishes a robust foundation for further investigations into the study of quantum systems and propels the convergence of quantum and artificial intelligence technology development for near-term devices.
	
	\section*{Acknowledgement} This work is support by GRF (grant no. 16305121). M.Y. and D.Z. are supported by National Key Research and Development Program of China (grant no. 2021YFA0718302 and no. 2021YFA1402104), the National Natural Science Foundation of China (grant no. 12075310), and the Strategic Priority Research Program of the Chinese Academy of Sciences (grant no. XDB28000000).
	
	\bibliographystyle{apsrev4-2} \bibliography{transformer.bbl}
	~\\
	
	\appendix
	\onecolumngrid
	\section{Transformer Model and Self-attention mechanism}
	In this section,we delve deeper into the specifics of the transformer model, a groundbreaking architecture that has transformed the landscape of natural language processing, machine translation, and many other fields. For physicists seeking to gain an in-depth understanding of this pioneering model, we provide a detailed description of its components, including the encoder, decoder, positional encoding, attention mechanism, and feedforward layers.
	
	The transformer model was introduced by Vaswani et al. in their 2017 paper, "Attention is All You Need" \cite{vaswani2017attention}. It is founded on the concept of self-attention, which enables the model to selectively focus on different parts of a sequence. In contrast to traditional recurrent neural networks (RNNs) and convolutional neural networks (CNNs), the transformer architecture achieves superior performance with significantly reduced training time by leveraging parallelization and self-attention mechanisms.
	
		\begin{figure}[H]
		\centering
		\includegraphics[width=\textwidth]{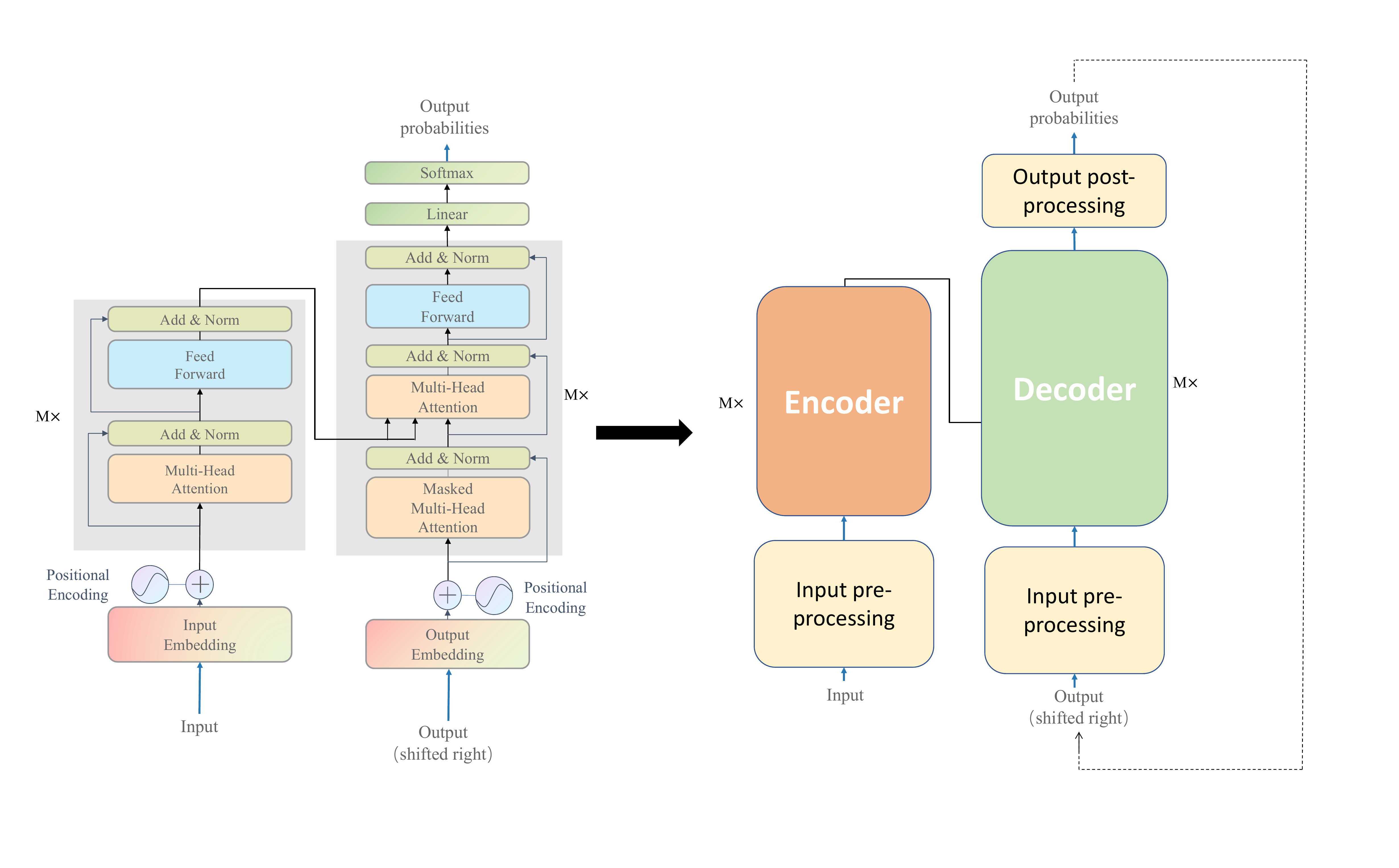}
		\caption{(left) The Transformer architecture. (right) An abstracted version of the same for better understanding.}
	\end{figure}
	
	The transformer architecture comprises two primary components: an encoder and a decoder. The encoder is responsible for processing the input sequence and generating a continuous representation, while the decoder generates the output sequence based on the encoder's representation. Each component is constructed from multiple layers, which are composed of several subcomponents. These subcomponents include the multi-head self-attention mechanism, the position-wise feedforward network, and the residual connections with layer normalization.
	
	The Algorithm~\ref{algo} illustrates the training procedure for the translation task using the transformer model, as described in the main text. The training dataset is denoted as $\mathcal{D}$, which contains individual tasks. For QST task, the input data consists of Hamiltonian parameters $\vec{x}$, while the output data comprises the measurement outcomes $\vec{b}$. The objective of the model is to accurately predict the conditional probability $P(\vec{b}|\vec{x})$. In the case of the Hamiltonian learning task, the input data is represented by the distribution of measurement outcomes $P(\vec{b}|\vec{x})$, and the output data consists of the Hamiltonian parameters $\vec{x}$. The model aims to predict the true value of the probability $P(\vec{x}|P(\vec{b}|\vec{x}))$.
	
		\begin{algorithm}[H]
		\caption{Training the Translator Model}
		\label{algo}
		\begin{algorithmic}[1]
			\Procedure{Train Translator}{$train\ data$, $transformer\ model$, $optimizer$, $loss\ function$, $num\ epochs$}
			\For{$epoch \gets 1$ \textbf{to} $num\ epochs$}
			\ForAll{$(input\ data, output\ data) \in train\ data$}
			\State $encoder\ input \gets \texttt{encoder\ pre-processing}(input\ data)$
			\State $encoder\ output \gets \texttt{encoder}(encoder\ input)$
			\State $decoder\ output \gets \texttt{decoder}(\texttt{decoder\ pre-processing}(output\ data),encoder\ output) $
			\State $probabilities \gets \texttt{output\ post-processing}(decoder\ output)$
			\State $loss \gets \texttt{loss\ function}(probabilities, output\ data)$
			\State $\texttt{optimizer.zero\ grad()}$
			\State $\texttt{loss.backward()}$
			\State $\texttt{optimizer.step()}$
			\EndFor
			\EndFor
			\EndProcedure
		\end{algorithmic}
	\end{algorithm}
	
	The encoder leverages an embedding layer to convert the input sequence into acquired vector representations of dimension $d_\text{model}$. In the decoder, a same embedding layer maps one-hot encoded input tokens to learned embeddings. In both the encoder and decoder, positional information is incorporated via a positional encoding layer, enabling the model to utilize both relative and absolute positions of tokens within a sequence. The positional encoding is crucial for the model to capture the sequential nature of the input data, as the self-attention mechanism is inherently permutation-invariant.
	
	Positional encodings are generated using sine and cosine functions of varying frequencies, dependent on the dimension and position $k$ in the sequence:
	\begin{equation}
		p e(k, 2 i)=\sin \left(\frac{k}{10000^{2 i / d_{\text {model }}}}\right), \quad p e(k, 2 i+1)=\cos \left(\frac{k}{10000^{2 i / d_{\text {model }}}}\right),
	\end{equation}
	where $i$ denotes the dimension. The sine and cosine functions ensure that the positional encoding remains differentiable and allows the model to learn and generalize to sequences of varying lengths. The embedded input tokens and positional encodings are summed before being passed to the subsequent layers.
	
	One of the most critical components of the transformer model is the attention mechanism. An attention function maps a query $Q$ and a collection of key-value pairs $K, V$ to an output, computed as a weighted sum with weights determined by the query and key. The transformer architecture employs a specialized attention function called scaled dot-product attention:
	\begin{equation}
		\operatorname{Attention}(Q, K, V)=\operatorname{softmax}\left(\frac{Q K^T}{\sqrt{d_k}}\right) V
	\end{equation}
	where $Q, K$, and $V$ are linear transformations of the input vectors:
	\begin{equation}
		Q=X W^Q, \quad K=X W^K, \quad V=X W^V.
	\end{equation}
	Here, $X \in \mathbb{R}^{n \times d_{\text {model }}}$ symbolizes the matrix of $n$ embedded input tokens with dimension $d_\text{model}$ and the projections are parameter matrices $W^Q \in \mathbb{R}^{d_{\text {model }} \times d_q}$, $W^K \in \mathbb{R}^{d_{\text {model }} \times d_k}$, $W^V \in \mathbb{R}^{d_{\text {model }} \times d_v}$.
	
	The scaled dot-product attention mechanism calculates the similarity between the query and the key, which is then used to determine the weight of each value in the output. This allows the model to selectively focus on the most relevant parts of the input sequence during processing.
	
	As delineated in~\cite{vaswani2017attention}, the transformer model adopts multi-head attention, in which the input vectors are linearly projected $n_h$ times into query, key, and value vectors, resulting in $n_h$ attention vectors. These vectors are subsequently concatenated and projected once more to produce the final output of the multi-head self-attention module:
\begin{equation}
	\begin{aligned}
		\operatorname{MultiHead}(Q, K, V) & =\operatorname{Concat}\left(\operatorname{head}_1, \ldots, \operatorname{head}_{\mathrm{n_h}}\right) W^O \\
		\text { where } \operatorname{head}_{\mathrm{i}} & =\operatorname{Attention}\left(Q W_i^Q, K W_i^K, V W_i^V\right).
	\end{aligned}
\end{equation}
	
	The multi-head attention mechanism allows the model to capture different aspects of the inputdata by focusing on various features simultaneously. This leads to a richer understanding of the relationships between different parts of the sequence, ultimately enhancing the model's performance.
	
	Following the multi-head attention layer, a position-wise feedforward network is employed, consisting of a fully connected neural network with two linear transformations and a ReLU activation applied discretely and identically to each position. This feedforward network is used to process the output of the multi-head attention layer and extract higher-level features from the input data.
	
	Each sublayer within the encoder and decoder components (i.e., self-attention or position-wise feedforward) incorporates a residual connection, which helps mitigate the vanishing gradient problem commonly encountered in deep neural networks. A residual connection computes the element-wise sum of the sublayer's input and output, allowing the gradient to flow more smoothly during backpropagation. Each residual connection is followed by layer normalization, a technique that normalizes the output across the layer to improve training stability and convergence.
	
	The encoder is composed of a stack of identical layers, each containing a multi-head self-attention sublayer followed by a position-wise feedforward network sublayer. Similarly, the decoder also consists of a stack of identical layers but with an additional multi-head attention sublayer inserted between the self-attention and feedforward sublayers. This additional attention sublayer in the decoder is responsible for attending to the output of the encoder, providing a bridge between the input and output sequences.
	
	The output of the final decoder layer is passed through a linear projection and a softmax layer, producing a probability distribution over the target vocabulary. During training, the model is optimized to minimize the cross-entropy loss between the predicted output and the ground truth target sequence.
	
	The transformer architecture is highly modular and can be easily scaled to accommodate larger datasets and more complex tasks. One such example is the BERT model \cite{devlin2018bert}, which builds upon the transformer architecture to create a bidirectional representation of the input text. BERT has achieved state-of-the-art performance on a wide range of natural language processing tasks, demonstrating the flexibility and power of the transformer model. Another notable example is the Generative Pre-trained Transformer (GPT) series \cite{radford2018improving,radford2019language,brown2020language,ouyang2022training}, which leverages a unidirectional transformer architecture and focuses on language modeling and generation. GPT models are pretrained on vast amounts of text data, enabling them to generate coherent and contextually relevant text based on a given input prompt. The GPT series has been applied to various tasks, such as machine translation, summarization, question-answering, and even conversational AI, showcasing the immense adaptability and potential of the transformer model in addressing diverse challenges in natural language processing and beyond.
	
	\section{Pauli-6 POVM}
	The single-qubit Pauli-6 POVM has six outcomes corresponding to sub-normalized rank-1 projections
	\begin{equation}
		\begin{aligned}
			\boldsymbol{M}_{\text {Pauli-6 }}=\{&M^{(0)}=\frac{1}{3} \times|0 \rangle\langle 0|, M^{(1)}= \frac{1}{3} \times|1\rangle \langle 1|, M^{(2)}=\frac{1}{3} \times |+\rangle \langle+ |,\\
			&M^{(3)}=\frac{1}{3} \times |-\rangle \langle-|, M^{(4)}=\frac{1}{3} \times| r \rangle \langle r |, M^{(5)}=\frac{1}{3} \times| l\rangle\langle l|\},
		\end{aligned}
	\end{equation}
	where $\{|+\rangle,|-\rangle\},\{|+i\rangle,|-i\rangle\}$ and $\{|0\rangle,|1\rangle\}$ are the eigenbases of the Pauli operators X, Y and Z, respectively.
	It is worth noting that each Pauli matrix, as well as the identity matrix, can be obtained from real linear combinations of the projections in $\mathcal{M}_{\text {Pauli-6 }}$. Therefore, the single-qubit Pauli-6 POVM, which is comprised of the POVM elements in $\mathcal{M}_{\text {Pauli-6 }}$, spans the space of $2\times2$ Hermitian matrices. By taking n-fold tensor products of the POVM elements in $\mathcal{M}_{\text {Pauli-6 }}$, the Pauli-6 POVM on n qubits is formed, and it is informationally complete.
	
	\section{CONSTRUCTING DENSITY MATRIX FROM THE PROBABILITY DISTRIBUTION}\label{sec:recover}
	In this section, we elucidate the methodology for transmuting a probability distribution $P(\vec{b})$ into a density matrix $\rho$. Owing to Born's rule, the probability distribution $P(\vec{b})$ spanning measurement outcomes $\vec{b} = \{b_1, b_2, \ldots, b_n\}$ on a quantum state $\rho$, characterized by $P(\vec{b}) \geq 0$ and $\sum_{\vec{b}} P(\vec{b}) = 1$, is furnished by the linear expression $P(\vec{b}) = \operatorname{Tr}\left[M^{(\vec{b})} \rho\right]$. The density matrix can be unambiguously deduced from the probability distribution of measurement outcomes. This relationship can be succinctly articulated when the overlap matrix $T$, comprising elements $T_{\vec{b}, \vec{b}^{\prime}} = \operatorname{Tr}\left[M^{(\vec{b})} M^{\left(\vec{b}^{\prime}\right)}\right]$, is invertible:
	\begin{equation}
		\rho=\sum_{\vec{b}, \vec{b}^{\prime}} P(\vec{b}) T_{\vec{b}, \vec{b}^{\prime}}^{-1} M^{\left(\vec{b}^{\prime}\right)}=\mathbb{E}_{\vec{b} \sim \boldsymbol{P}}\left(\sum_{\vec{b}^{\prime}} T_{\vec{b}, \vec{b}^{\prime}}^{-1} M^{\left(\vec{b}^{\prime}\right)}\right),
	\end{equation}
	where $\mathbb{E}_{\vec{b} \sim \boldsymbol{P}}$ denotes the expectation value over $\vec{b}$ distributed in accordance with $\boldsymbol{P}$.
	
	\section{CLASSICAL SHADOWS}\label{sec:shadow}
	
	Here we introduce the method of Shadow Tomography~\cite{huang2020predicting,huang2021efficient}. First, apply a random unitary to rotate the state ($\rho \mapsto U \rho U^{\dagger}$) and perform a computational-basis measurement. Then, after the measurement, they apply the inverse of $U$ to the resulting computational basis state. This procedure collapses $\rho$ to a snapshot $U^{\dagger}|\hat{b}\rangle\langle \hat{b}|U$, producing a quantum channel $\mathcal{M}$, which depends on the ensemble of (random) unitary transformations.
	
	If the collection of unitaries is defined to be tomographically complete, namely, if the condition i.e. for each $\sigma \neq \rho$, there exist $U \in \mathcal{U}$ and b such that $\left\langle b\left|U \sigma U^{\dagger}\right| b\right\rangle \neq  \left\langle b\left|U \rho U^{\dagger}\right| b\right\rangle $ is met, then $\mathcal{U}$ — viewed as a linear map — has a unique inverse $\mathcal{U}^{-1}$. As~\cite{huang2020predicting} set
	
	\begin{equation}
		\hat{\rho}=\mathcal{M}^{-1}\left(U^{\dagger}|\hat{b}\rangle\langle\hat{b}| U\right).
	\end{equation}

	For local measurements, the inverse channel for the n-qubit system can be written as
	\begin{equation}
		\mathcal{M}_n^{-1}=\bigotimes_{j=1}^n \mathcal{M}_1^{-1}.
	\end{equation}

We can now reformulate the shadows with our overcomplete POVM set and its corresponding channel. For Pauli-6 POVM, we will get
\begin{equation}
	\hat{\rho}=\bigotimes_{j=1}^n \mathcal{M}_1^{-1}\left(\left|\psi_{a, j}\right\rangle\left\langle\psi_{a, j}\right|\right),
\end{equation}
where $\mathcal{M}_1^{-1}(X)=3 X-\operatorname{tr}(X) \mathbf{I}$ and $X=\left(x_0 \mathbf{I}+\vec{r} \cdot \vec{\sigma}\right)$ is a 2 dimensional (single-qubit) quantum operation with the Bloch representation $\rho=\frac{1}{2}(\mathbf{I}+\vec{r} \cdot \vec{\sigma})$. Note that
the $2^n \times 2^n$ matrix $\rho$ need not be constructed explicitly. We just need to store $|\psi_{a, j}\rangle$ for each qubit j.

\section{Kernel Methods and the Neural Tangent Kernel}

	Kernel methods constitute a class of classical machine learning algorithms that employ a kernel function to implicitly transform input data into a higher-dimensional representation. The fundamental premise is that mapping the data to this higher-dimensional space facilitates the discovery of linear patterns within the data, which can ultimately be harnessed for tasks such as classification or regression.
	
	The Neural Tangent Kernel (NTK) is a kernel method explicitly tailored for neural networks. It is derived from the tangent kernel, which represents a linear approximation of the neural network's function surrounding the current parameters~\cite{jacot2018ntk}. The NTK encapsulates the behavior of the neural network during training, enabling the examination of training dynamics and the formulation of predictions using the kernel method. The NTK is defined as the inner product between two gradient vectors, which can be expressed as:
	\begin{equation}
		K(\Vec{x}, \Vec{x}')=
		\sum_{m=1}^M
		\frac{\partial h(\Vec{x}, \Vec{\theta})}{\partial \theta_m}
		\cdot
		\frac{\partial h(\Vec{x}', \Vec{\theta})}{\partial \theta_m}
	\end{equation}
	where $h(\Vec{x}, \Vec{\theta})$ is the output of the neural network with input $\Vec{x}$ and parameters $\Vec{\theta}$ and $M$ represents the number of parameters in the neural network.
	
	Given $N$ training data points  $\Vec{x}^{(1)},\ldots,\Vec{x}^{(N)}$ and corresponding target values $y^{(1)},\ldots,y^{(N)}$, the NTK regression model can be written as follows:
	\begin{equation}
		f(\Vec{x})=a_0+\sum_{n=1}^N a_n K(\Vec{x}^{(n)},\Vec{x}),
	\end{equation}
	where $a_0,\ldots,a_{N}$ are the regression coefficients. To obtain the coefficients $\Vec{a}$, we can solve the following optimization problem:
	\begin{equation}
		\min_{\Vec{a}}\frac{1}{N}\sum_{n=1}^N(y_n-f(\Vec{x}^{(n)}))^2+\lambda\sum_{n=0}^N a_n^2,
	\end{equation}
	where $\lambda$ is a regularization parameter that controls the trade-off between fitting the data and preventing over-fitting. This optimization problem can be solved using standard techniques such as ridge regression. To make a prediction for a new input $\Vec{x}$, we first compute the kernel function between $\Vec{x}$ and each training point $\Vec{x}^{(n)}$, and then use these kernel values along with the regression coefficients $\Vec{a}$ to compute the predicted output $f(\Vec{x})$. 
	
	In this study, the NTK method is employed to make predictions for comparison with our translator model. For instance, we apply the NTK method to predict the ground state properties of 2D antiferromagnetic random Heisenberg models. Specifically, to predict the expectation value of the correlation function $\langle C_{i,j} \rangle$ for each site pair $i,j$ in the Heisenberg model, we train a distinct NTK model $f_{i,j}$ by adjusting the regularization parameter $\lambda_{i,j}$.
	
	For each pair $i,j$, we use a training dataset $\mathcal{D}_{i,j}=\left(\langle C_{i,j} \rangle^{(1)}, \vec{x}^{(1)}\right), \ldots,\left(\langle C_{i,j} \rangle^{(N)}, \vec{x}^{(N)}\right)$ comprising $N$ samples, where $\vec{x}^{(n)}$ represents the Hamiltonian parameters for the $n$-th training sample. To ensure consistency with our translation model, the training data here is collected based on the classical shadow formalism. For a given set of Hamiltonian parameters $\vec{x}$, the corresponding ground state $\rho(\vec{x})$ can be prepared using the specified Hamiltonian $H(\vec{x})$. Subsequently, the probability distribution of measurement $P(\vec{b}|\vec{x})$ can be obtained using the classical shadow technique with the ground state $\rho(\vec{x})$. We remark that the dataset $\mathcal{D}=\left(\vec{b}^{(1)}, \vec{x}^{(1)}\right), \ldots,\left(\vec{b}^{(N)}, \vec{x}^{(N)}\right)$, which consists of $N$ samples, serves as the training dataset for our translation model. The classical shadow representation for ground state $\hat{\rho}(\vec{x})$ then can be reconstructed from the probability distribution of measurement $P(\vec{b}|\vec{x})$. For each site $i$, $j$ in the Heisenberg model, the expected correlation function $\langle C_{i,j} \rangle$ can be obtained by $\langle C_{i,j} \rangle = \operatorname{Tr}\left[C_{i,j}\rho(\vec{x})\right]$, which serves as the target value for the NTK model $f_{i,j}$.
	
	For the ground state property prediction task, it is crucial to recognize that the NTK model learns a classical-classical mapping, in which the Hamiltonian parameters are mapped to the expectation value of a specific observable~\cite{Huang2022}. This suggests that the NTK model does not rely on quantum data during training. To predict the expectation value of a new observable of interest, a new corresponding training dataset must be generated, and a new NTK model needs to be trained following the aforementioned procedure. In contrast, our translation model leverages quantum measurement data to learn the relationship between the Hamiltonian parameters and the corresponding ground state, and it does not necessitate the collection of a new training dataset for a new observable of interest.

\end{document}